\input epsf
%
%
%
\def\unredoffs{} 

%
%
%
%
\newbox\leftpage \newdimen\fullhsize \newdimen\hstitle \newdimen\hsbody
\tolerance=1000\hfuzz=2pt
\catcode`\@=11 
%
\magnification=1200\unredoffs\baselineskip=16pt plus 2pt minus 1pt
\hsbody=\hsize \hstitle=\hsize 
%
%
%
\newcount\yearltd\yearltd=\year\advance\yearltd by -1900

%
%

\def\draftmode{\message{ DRAFTMODE }\def\draftdate{{\rm preliminary draft:
\number\month/\number\day/\number\yearltd\ \ \hourmin}}%
\headline={\hfil\draftdate}\writelabels\baselineskip=20pt plus 2pt minus 2pt
 {\count255=\time\divide\count255 by 60 \xdef\hourmin{\number\count255}
  \multiply\count255 by-60\advance\count255 by\time
  \xdef\hourmin{\hourmin:\ifnum\count255<10 0\fi\the\count255}}}
\def\nolabels{\def\wrlabeL##1{}\def\eqlabeL##1{}\def\reflabeL##1{}}
\def\writelabels{\def\wrlabeL##1{\leavevmode\vadjust{\rlap{\smash%
{\line{{\escapechar=` \hfill\rlap{\sevenrm\hskip.03in\string##1}}}}}}}%
\def\eqlabeL##1{{\escapechar-1\rlap{\sevenrm\hskip.05in\string##1}}}%
\def\reflabeL##1{\noexpand\llap{\noexpand\sevenrm\string\string\string##1}}}
\nolabels
%
\global\newcount\secno \global\secno=0
\global\newcount\meqno \global\meqno=1
\def\newsec#1{\global\advance\secno by1\message{(\the\secno. #1)}
\global\subsecno=0\eqnres@t\noindent{\bf\the\secno. #1}
\writetoca{{\secsym} {#1}}\par\nobreak\medskip\nobreak}
\def\eqnres@t{\xdef\secsym{\the\secno.}\global\meqno=1\bigbreak\bigskip}
\def\sequentialequations{\def\eqnres@t{\bigbreak}}\xdef\secsym{}
\global\newcount\subsecno \global\subsecno=0
\def\subsec#1{\global\advance\subsecno by1\message{(\secsym\the\subsecno. #1)}
\ifnum\lastpenalty>9000\else\bigbreak\fi
\noindent{\it\secsym\the\subsecno. #1}\writetoca{\string\quad
{\secsym\the\subsecno.} {#1}}\par\nobreak\medskip\nobreak}
\def\appendix#1#2{\global\meqno=1\global\subsecno=0\xdef\secsym{\hbox{#1.}}
\bigbreak\bigskip\noindent{\bf Appendix #1 #2}\message{(#1 #2)}
\writetoca{Appendix {#1} {#2}}\par\nobreak\medskip\nobreak}
%
%
\def\eqnn#1{\xdef #1{(\secsym\the\meqno)}\writedef{#1\leftbracket#1}%
\global\advance\meqno by1\wrlabeL#1}
\def\eqna#1{\xdef #1##1{\hbox{$(\secsym\the\meqno##1)$}}
\writedef{#1\numbersign1\leftbracket#1{\numbersign1}}%
\global\advance\meqno by1\wrlabeL{#1$\{\}$}}
\def\eqn#1#2{\xdef #1{(\secsym\the\meqno)}\writedef{#1\leftbracket#1}%
\global\advance\meqno by1$$#2\eqno#1\eqlabeL#1$$}
%
\newskip\footskip\footskip14pt plus 1pt minus 1pt 
\def\footnotefont{\ninepoint}\def\f@t#1{\footnotefont #1\@foot}
\def\f@@t{\baselineskip\footskip\bgroup\footnotefont\aftergroup\@foot\let\next}
\setbox\strutbox=\hbox{\vrule height9.5pt depth4.5pt width0pt}
\global\newcount\ftno \global\ftno=0
\def\foot{\global\advance\ftno by1\footnote{$^{\the\ftno}$}}
%
\newwrite\ftfile
\def\footend{\def\foot{\global\advance\ftno by1\chardef\wfile=\ftfile
$^{\the\ftno}$\ifnum\ftno=1\immediate\openout\ftfile=foots.tmp\fi%
\immediate\write\ftfile{\noexpand\smallskip%
\noexpand\item{f\the\ftno:\ }\pctsign}\findarg}%
\def\footatend{\vfill\eject\immediate\closeout\ftfile{\parindent=20pt
\centerline{\bf Footnotes}\nobreak\bigskip\input foots.tmp }}}
\def\footatend{}
%
%
\global\newcount\refno \global\refno=1
\newwrite\rfile
\def\ref{[\the\refno]\nref}
\def\nref#1{\xdef#1{[\the\refno]}\writedef{#1\leftbracket#1}%
\ifnum\refno=1\immediate\openout\rfile=refs.tmp\fi
\global\advance\refno by1\chardef\wfile=\rfile\immediate
\write\rfile{\noexpand\item{#1\ }\reflabeL{#1\hskip.31in}\pctsign}\findarg}
\def\findarg#1#{\begingroup\obeylines\newlinechar=`\^^M\pass@rg}
{\obeylines\gdef\pass@rg#1{\writ@line\relax #1^^M\hbox{}^^M}%
\gdef\writ@line#1^^M{\expandafter\toks0\expandafter{\striprel@x #1}%
\edef\next{\the\toks0}\ifx\next\em@rk\let\next=\endgroup\else\ifx\next\empty%
\else\immediate\write\wfile{\the\toks0}\fi\let\next=\writ@line\fi\next\relax}}
\def\striprel@x#1{} \def\em@rk{\hbox{}}
\def\lref{\begingroup\obeylines\lr@f}
\def\lr@f#1#2{\gdef#1{\ref#1{#2}}\endgroup\unskip}
\def\semi{;\hfil\break}
\def\addref#1{\immediate\write\rfile{\noexpand\item{}#1}} 
\def\startrefs#1{\immediate\openout\rfile=refs.tmp\refno=#1}
\def\xref{\expandafter\xr@f}\def\xr@f[#1]{#1}
\def\refs#1{\count255=1[\r@fs #1{\hbox{}}]}
\def\r@fs#1{\ifx\und@fined#1\message{reflabel \string#1 is undefined.}%
\nref#1{need to supply reference \string#1.}\fi%
\vphantom{\hphantom{#1}}\edef\next{#1}\ifx\next\em@rk\def\next{}%
\else\ifx\next#1\ifodd\count255\relax\xref#1\count255=0\fi%
\else#1\count255=1\fi\let\next=\r@fs\fi\next}
%

%
\newwrite\ffile\global\newcount\figno \global\figno=1
\def\fig{fig.~\the\figno\nfig}
\def\nfig#1{\xdef#1{fig.~\the\figno}%
\writedef{#1\leftbracket fig.\noexpand~\the\figno}%
\ifnum\figno=1\immediate\openout\ffile=figs.tmp\fi\chardef\wfile=\ffile%
\immediate\write\ffile{\noexpand\medskip\noexpand\item{Fig.\ \the\figno. }
\reflabeL{#1\hskip.55in}\pctsign}\global\advance\figno by1\findarg}
\def\xfig{\expandafter\xf@g}\def\xf@g fig.\penalty\@M\ {}
\def\figs#1{figs.~\f@gs #1{\hbox{}}}
\def\f@gs#1{\edef\next{#1}\ifx\next\em@rk\def\next{}\else
\ifx\next#1\xfig #1\else#1\fi\let\next=\f@gs\fi\next}
\newwrite\lfile
{\escapechar-1\xdef\pctsign{\string\%}\xdef\leftbracket{\string\{}
\xdef\rightbracket{\string\}}\xdef\numbersign{\string\#}}

\def\writestop{\def\writestoppt{\immediate\write\lfile{\string\pageno%
\the\pageno\string\startrefs\leftbracket\the\refno\rightbracket%
\string\def\string\secsym\leftbracket\secsym\rightbracket%
\string\secno\the\secno\string\meqno\the\meqno}\immediate\closeout\lfile}}
\def\writestoppt{}\def\writedef#1{}
\def\seclab#1{\xdef #1{\the\secno}\writedef{#1\leftbracket#1}\wrlabeL{#1=#1}}
\def\subseclab#1{\xdef #1{\secsym\the\subsecno}%
\writedef{#1\leftbracket#1}\wrlabeL{#1=#1}}
\newwrite\tfile \def\writetoca#1{}
\def\leaderfill{\leaders\hbox to 1em{\hss.\hss}\hfill}
\def\writetoc{\immediate\openout\tfile=toc.tmp
   \def\writetoca##1{{\edef\next{\write\tfile{\noindent ##1
   \string\leaderfill {\noexpand\number\pageno} \par}}\next}}}

%
\catcode`\@=12 
%
\edef\tfontsize{\ifx\answ\bigans scaled\magstep3\else scaled\magstep4\fi}
\font\titlerm=cmr10 \tfontsize \font\titlerms=cmr7 \tfontsize
\font\titlermss=cmr5 \tfontsize \font\titlei=cmmi10 \tfontsize
\font\titleis=cmmi7 \tfontsize \font\titleiss=cmmi5 \tfontsize
\font\titlesy=cmsy10 \tfontsize \font\titlesys=cmsy7 \tfontsize
\font\titlesyss=cmsy5 \tfontsize \font\titleit=cmti10 \tfontsize
\skewchar\titlei='177 \skewchar\titleis='177 \skewchar\titleiss='177
\skewchar\titlesy='60 \skewchar\titlesys='60 \skewchar\titlesyss='60
\def\titlefont{\def\rm{\fam0\titlerm}
\textfont0=\titlerm \scriptfont0=\titlerms \scriptscriptfont0=\titlermss
\textfont1=\titlei \scriptfont1=\titleis \scriptscriptfont1=\titleiss
\textfont2=\titlesy \scriptfont2=\titlesys \scriptscriptfont2=\titlesyss
\textfont\itfam=\titleit \def\it{\fam\itfam\titleit}\rm}
 \ifx\answ\bigans\else scaled\magstep1\fi
\ifx\answ\bigans\else

 \font\absi=cmmi10 scaled\magstep1
\font\absis=cmmi7 scaled\magstep1 \font\absiss=cmmi5 scaled\magstep1
\font\abssy=cmsy10 scaled\magstep1 \font\abssys=cmsy7 scaled\magstep1
\font\abssyss=cmsy5 scaled\magstep1 
\skewchar\absi='177 \skewchar\absis='177 \skewchar\absiss='177
\skewchar\abssy='60 \skewchar\abssys='60 \skewchar\abssyss='60
\fi
\font\ninerm=cmr9 \font\sixrm=cmr6 \font\ninei=cmmi9 \font\sixi=cmmi6
\font\ninesy=cmsy9 \font\sixsy=cmsy6 \font\ninebf=cmbx9
\font\nineit=cmti9 \font\ninesl=cmsl9 \skewchar\ninei='177
\skewchar\sixi='177 \skewchar\ninesy='60 \skewchar\sixsy='60
\def\ninepoint{\def\rm{\fam0\ninerm}
\textfont0=\ninerm \scriptfont0=\sixrm \scriptscriptfont0=\fiverm
\textfont1=\ninei \scriptfont1=\sixi \scriptscriptfont1=\fivei
\textfont2=\ninesy \scriptfont2=\sixsy \scriptscriptfont2=\fivesy
\textfont\itfam=\ninei \def\it{\fam\itfam\nineit}\def\sl{\fam\slfam\ninesl}%
\textfont\bffam=\ninebf \def\bf{\fam\bffam\ninebf}\rm}
%
%
\def\noblackbox{\overfullrule=0pt}
\hyphenation{anom-aly anom-alies coun-ter-term coun-ter-terms}
\def\inv{^{\raise.15ex\hbox{${\scriptscriptstyle -}$}\kern-.05em 1}}

\def\Dsl{\,\raise.15ex\hbox{/}\mkern-13.5mu D} 
\def\dsl{\raise.15ex\hbox{/}\kern-.57em\partial}

 \def\Tr{{\rm Tr}}

\def\lspace{\ifx\answ\bigans{}\else\qquad\fi}
\def\lbspace{\ifx\answ\bigans{}\else\hskip-.2in\fi} 
\def\boxeqn#1{\vcenter{\vbox{\hrule\hbox{\vrule\kern3pt\vbox{\kern3pt
        \hbox{${\displaystyle #1}$}\kern3pt}\kern3pt\vrule}\hrule}}}
\def\mbox#1#2{\vcenter{\hrule \hbox{\vrule height#2in
                \kern#1in \vrule} \hrule}}  
%

\def\darr#1{\raise1.5ex\hbox{$\leftrightarrow$}\mkern-16.5mu #1}

\def\half{{\textstyle{1\over2}}} 
\def\roughly#1{\raise.3ex\hbox{$#1$\kern-.75em\lower1ex\hbox{$\sim$}}}
\hyphenation{Mar-ti-nel-li}

\def\Re{\,\hbox{Re}\,}

\def\Tr{\,{\hbox{Tr}}\,}

\def\1{\;1\!\!\!\! 1\;}

\def\ie{{\it i.e.}}

\def\etal{{\it et al.}}

\def\frac#1#2{{{#1}\over {#2}}}
\def\half{\hbox{${1\over 2}$}}

\def\smallfrac#1#2{\hbox{${{#1}\over {#2}}$}}

\def\Tr{{\rm Tr}}

\def\MS{\hbox{$\overline{\rm MS}$}}
\def\ms{\hbox{$\overline{\scriptstyle\rm MS}$}}
\def\QMS{Q$_0$\MS}

\catcode`@=11 
\def\slash#1{\mathord{\mathpalette\c@ncel#1}}
 \def\c@ncel#1#2{\ooalign{$\hfil#1\mkern1mu/\hfil$\crcr$#1#2$}}
\def\lsim{\mathrel{\mathpalette\@versim<}}
\def\gsim{\mathrel{\mathpalette\@versim>}}
 \def\@versim#1#2{\lower0.2ex\vbox{\baselineskip\z@skip\lineskip\z@skip
       \lineskiplimit\z@\ialign{$\m@th#1\hfil##$\crcr#2\crcr\sim\crcr}}}
\catcode`@=12 

\def\PR{{\it Phys.~Rev.~}}
\def\PRL{{\it Phys.~Rev.~Lett.~}}
\def\NP{{\it Nucl.~Phys.~}}

\def\PL{{\it Phys.~Lett.~}}

\def\SJNP{{\it Sov.~Jour.~Nucl.~Phys.~}}
\def\SPJETP{{\it Sov.~Phys.~J.E.T.P.~}}

\def\JHEP{{\it Jour.~High~Energy~Phys.~}}
\def\vol#1{{\bf #1}}\def\vyp#1#2#3{\vol{#1} (#2) #3}

\def\as{\alpha_s}
\def\ahat{\hat\as}

\noblackbox
\pageno=0\nopagenumbers\tolerance=10000\hfuzz=5pt
\baselineskip 12pt
\line{\hfill CERN-PH-TH/2008-014}
\line{\hfill Edinburgh 2007/49}
\line{\hfill IFUM-905-FT}
\line{\hfill RM3-TH/08-1}
\vskip 12pt
\centerline{\titlefont Small $x$ Resummation with Quarks:}
\centerline{\titlefont Deep-Inelastic Scattering}
\vskip 36pt\centerline{Guido~Altarelli,$^{(a,\>b)}$
Richard D.~Ball$^{(c,\>b)}$ and Stefano Forte$^{(d)}$}
\vskip 12pt
\centerline{\it ${}^{(a)}$ Dipartimento di Fisica ``E.Amaldi'', 
Universit\`a Roma Tre}
\centerline{\it INFN, Sezione di Roma Tre}
\centerline{\it Via della Vasca Navale 84, I--00146 Roma, Italy}
\vskip 6pt
\centerline{\it ${}^{(b)}$ CERN, Department of Physics, Theory Division}
\centerline{\it CH-1211 Gen\`eve 23, Switzerland}
\vskip 6pt
\centerline{\it ${}^{(c)}$School of Physics, University of Edinburgh}
\centerline{\it  Edinburgh EH9 3JZ, Scotland}
\vskip 6pt
\centerline {\it ${}^{(d)}$Dipartimento di  Fisica, Universit\`a di
Milano and}
\centerline{\it INFN, Sezione di Milano, Via Celoria 16, I-20133 Milan, Italy}
\vskip 20pt
\centerline{\bf Abstract}
{\narrower\baselineskip 10pt
\medskip\noindent 
We extend our previous results on  small- $x$ resummation 
in the pure Yang--Mills
theory to full QCD with $n_f$ quark flavours, with
a resummed two-by-two matrix of resummed quark and gluon splitting functions.
We also construct the corresponding deep--inelastic
coefficient functions, and show how
these can be combined with parton densities to give  
fully resummed deep--inelastic structure
functions $F_2$ and $F_L$  at
the next-to-leading logarithmic level. We discuss how this resummation can 
be performed in different factorization schemes, including
the commonly used \MS\ scheme. We study the importance of the resummation 
effects by comparison with fixed-order perturbative results, and we
discuss the corresponding
renormalization and factorization scale variation uncertainties. 
We find that for $x$ below $10^{-2}$ the resummation effects are comparable in 
size to the fixed order NNLO
corrections, but differ in shape. We finally discuss the 
phenomenological impact
of the small--$x$ resummation, specifically in the extraction of
parton distribution from present day experiments and their
extrapolation to the kinematics relevant for future colliders such as
the LHC.}
\vfill
\line{CERN-PH-TH/2008-014\hfill }
\line{January 2008\hfill}
\eject \footline={\hss\tenrm\folio\hss}
\lref\ciafqz{
  M.~Ciafaloni,  
  \PL\vyp{B356}{1995}{74}\semi
 M.~Ciafaloni and D.~Colferai,
  \JHEP\vyp{0509}{2005}{069}\semi
 S.~Marzani, R.~D.~Ball, P.~Falgari and S.~Forte,
  \NP\vyp{B783}{2007}{143}.
}
\lref\be{ R.~D.~Ball and R.~K.~Ellis,
  \JHEP\vyp{0105}{2001}{053}.
}

\lref\cfp{G.~Curci, W.~Furma\'nski and R.~Petronzio,
\NP\vyp{B175}{1980}{27} 
}
\lref\liprun{ L.~N.~Lipatov,
  \SPJETP\vyp{63}{1986}{904}
  [Zh.\ Eksp.\ Teor.\ Fiz.\ \vyp{90}{1986}{1536}].
}
\lref\mf{
S.~Forte and R.~D.~Ball,
AIP Conf.\ Proc.\ \vyp{602}{2001}{60}, 
{\tt hep-ph/0109235.}
}
\lref\summ{ R.~D.~Ball and S.~Forte,
  \PL\vyp{B351}{1995}{313}.
}
\lref\afp{ R.~D.~Ball and S.~Forte,
  \PL\vyp{B405}{1997}{317}.
}
\lref\salamrev{  G.~P.~Salam,
  {\tt hep-ph/0607153.}, invited talk at DIS2006 (Tsukuba), Apr 2006.
}
\lref\newa{A.~Gehrmann-De~Ridder, T.~Gehrmann, E.W.N.~Glover and  
G.~Heinrich, 
\PRL\vyp{99}{2007}{132002};
\JHEP\vyp{0712}{2007}{094}.
}
\lref\newb{P.A.~Baikov, K.G.~Chetyrkin, J.H.~Kuhn, 
{\tt arXiv:0801.1821 [hep-ph]}.}

\lref\nnlo{ S.~Moch, J.~A.~M.~Vermaseren and A.~Vogt,
  \NP\vyp{B691}{2004}{129}\semi
 A.~Vogt, S.~Moch and J.~A.~M.~Vermaseren,
  \NP\vyp{B688}{2004}{101}.
}

\lref\bassetto{ A.~Bassetto, M.~Ciafaloni and G.~Marchesini,
  Phys.\ Rept.\  \vyp{100}{1983}{201}. 
}
\lref\bfkl{L.N.~Lipatov,
\SJNP\vyp{23}{1976}{338}\semi 
 V.S.~Fadin, E.A.~Kuraev and L.N.~Lipatov,
\PL\vyp{60B}{1975}{50}; 
 {\it Sov. Phys. JETP~}\vyp{44}{1976}{443}; 
\vyp{45}{1977}{199}\semi 
 Y.Y.~Balitski and L.N.Lipatov,
\SJNP\vyp{28}{1978}{822}.} 

\lref\jar{T.~Jaroszewicz,
\PL\vyp{B116}{1982}{291}.}
\lref\sxap{R.~D.~Ball and S.~Forte,
\PL\vyp{B465}{1999}{271}.}
\lref\sxres{G. Altarelli, R.~D. Ball and S. Forte,
\NP{\bf B575}, 313 (2000);  
see also {\tt hep-ph/0001157}.
}
\lref\rcdual{R.~D.~Ball and S.~Forte,
\NP\vyp{B742}{2006}{158}.
}
\lref\sxphen{G. Altarelli, R.~D.~Ball and S. Forte,
\NP\vyp{B599}{2001}{383};  
see also {\tt hep-ph/0104246}.}  
\lref\sxrun{
G.~Altarelli, R.~D.~Ball and S.~Forte,
\NP\vyp{B621}{2002}{359}.
}
\lref\ciafresa{
 M.~Ciafaloni, D.~Colferai and G.~P.~Salam,
  \PR\vyp{D60}{1999}{114036};
\JHEP\vyp{0007}{2000}{054}.
}
\lref\ciafresb{M.~Ciafaloni, D.~Colferai, G.~P.~Salam and A.~M.~Stasto,
  \PR\vyp{D66}{2002}{054014}; 
  \PL\vyp{B576}{2003}{143};
  \PR\vyp{D68}{2003}{114003},
see also the Proceedings of the HERA-LHC workshop (2005), in press.
}
\lref\ciafdip{M.~Ciafaloni, D.~Colferai, G.~P.~Salam and A.~M.~Stasto,
  \PL\vyp{B587}{2004}{87}; see also
 G.~P.~Salam,
{\tt hep-ph/0501097}.
}
\lref\runph{G.~Altarelli, R.~D.~Ball and S.~Forte,
\NP\vyp{B674}{2003}{459};
see also   
 {\tt hep-ph/0310016.}
}
\lref\sxsym{
  G.~Altarelli, R.~D.~Ball and S.~Forte,
  \NP\vyp{B742}{2006}{1},
see also 
  S.~Forte, G.~Altarelli and R.~D.~Ball,
  {\tt hep-ph/0606323.}
}
\lref\heralhc{ M.~Dittmar {\it et al.},
  {\tt hep-ph/0511119.}
}
\lref\salam{G.~Salam, \JHEP\vyp{9807}{1998}{19}.}
\lref\ciaf{M.~Ciafaloni and D.~Colferai,
\PL\vyp{B452}{1999}{372}. 
}

\lref\ch{
S.~Catani and F.~Hautmann,
\PL\vyp{B315}{1993}{157}; 
\NP\vyp{B427}{1994}{475}. 
}
\lref\hefac{S.~Catani, M.~Ciafaloni and F.~Hautmann,
\PL\vyp{B242}{1990}{97};
  \NP\vyp{B366}{1991}{135}\semi
J.~C.~Collins and R.~K.~Ellis,
  \NP\vyp{B360}{1991}{3}.
}
\lref\matevol{ M.~Ciafaloni, D.~Colferai, G.~P.~Salam and A.~M.~Stasto,
  {\tt arXiv:0707.1453 [hep-ph]}.
}
\lref\ballhad{ R.~D.~Ball,
  {\tt arXiv:0708.1277 [hep-ph]}, to be published in \NP B.
}
\lref\mom{ R.~D.~Ball and S.~Forte,
  \PL\vyp{B359}{1995}{362}.
}
\lref\lkns{R.~Kirschner and L.~N.~Lipatov,
  \NP\vyp{B213}{1983}{122}.
}
\lref\thorne{R.S.~Thorne, 
\PR\vyp{D64}{2001}{074005}\semi
C.D.~White and R.S.~Thorne, 
\PR\vyp{D75}{2007}{034005}.
}
\newsec{Introduction}
\noindent 
The main motivation behind the recent progress in higher order
calculations in perturbative QCD is the need of accurate phenomenology
at hadron colliders, and in particular the LHC. Therefore, the
interest of such calculations lies mostly in their ability to actually
lead to an improvement in the accuracy of 
theoretical predictions of measurable processes. The frontier of
present-day perturbative calculations is the next-to-next-to leading
(NNLO) order, thanks to the recent determination of three--loop splitting
functions~\nnlo\ as well as the hard partonic cross sections for
several processes~\salamrev\ (for
recent results see, for example, \refs{\newa,\newb}). 
However, perturbative evolution at NNLO is
unstable in the high energy (small $x$ limit): the size of the NNLO
corrections diverges as $x\to 0$ at fixed scale. 

The fact that at high energy  corrections to perturbative
evolution and to hard cross sections are potentially large  has been known for 
a long time (see {\it e.g.} Ref.~\bassetto). However, even though the study of the
leading high energy corrections~\bfkl\ and their inclusion in
perturbative  anomalous dimensions~\jar\ has a rather long history, it is only 
over the last few years that a fully resummed approach to
perturbative evolution has been constructed ~\refs{\sxap,\ciafresb}, 
mostly stimulated by the
availability of deep--inelastic data at very high energy from the HERA
collider~\heralhc. In fact, it turns out that even though at fixed
perturbative order 
corrections are very large at small
$x$, their full resummation leads to a considerable softening of small
$x$ terms, consistent with the fact that the data do not show 
any large departure from next-to-leading (NLO) order predictions.  
However, in order to obtain the resummed results one must include several classes
of subleading terms, motivated by various
physical constraints, such as momentum conservation, renormalization
group invariance and gluon exchange symmetry. The existing approaches
to this resummation, as discussed respectively in
refs.~\refs{\sxap,\sxres\sxphen\sxrun\runph{--}\sxsym}
and~\refs{\ciafresb,\salam\ciaf\ciafresa{--}\ciafdip}
(see also ref. \thorne)
though rather different in many technical
respects, are essentially based on the same physical input and
yield results which agree with each other within the expected
theoretical uncertainty.

Because the leading high energy corrections are dominated by gluon
exchange, the resummation is most easily performed in the pure Yang-Mills
theory, and indeed the fully resummed results for perturbative
evolution of refs.~\refs{\sxap,\ciafresb,\sxres\sxphen\sxrun\runph\sxsym\salam\ciaf\ciafresa{--}\ciafdip} have been obtained with
$n_f=0$. In order to actually get predictions for (flavour singlet) physical
observables, one needs to combine three ingredients: 
the eigenvalues of the
singlet quark and gluon anomalous dimension matrix,
the resummed partonic cross sections (coefficient functions) for the
relevant physical process in some factorization scheme, and the linear 
transformation which relates
the eigenvectors of the evolution matrix to the singlet quark and the
gluon in the same factorization scheme.
The first ingredient is readily available: because the effect of
one of the two evolution eigenvalues  is
suppressed by a power of $x$ at small $x$, it is enough to generalize the results of
refs.~\refs{\sxap-\ciafdip} for the
``large'' eigenvalue to the case $n_f\not=0$. The second ingredient is 
also available, at least
for a small number of processes~\refs{\hefac\ch{--}\be} for which
the high--energy resummation of partonic cross
sections has been performed.
In particular,
deep--inelastic coefficient functions have been determined in the \MS\
and DIS (and related) schemes in ref.~\ch. However, it is nontrivial
to combine resummed coefficient functions and parton distributions at
the running coupling level: the way to do this has only been recently
developed in ref.~\ballhad.
Hence, in order to obtain complete results what
we need is to construct the 
evolution eigenvectors in terms of quarks and gluons
at the resummed level 
in these schemes. This can be done by extending to the running coupling case
the general technique for the construction of
resummed factorization schemes which was developed in
refs.~\refs{\sxphen,\mom}.

In this paper, we will present a construction of resummed physical
observables in the \MS\ and related schemes, based on the
approach to resummation of refs.~\refs{\sxap,\sxres-\sxsym}, using the
strategy that we just outlined, and apply it to the specific case of
deep--inelastic scattering. We start by summarising the results on small $x$ 
resummation of singlet evolution developed in previous papers for the pure 
Yang-Mills theory.  We then construct the two
eigenvalues of the anomalous dimension matrix when $n_f\not=0$, by a
suitable generalization of the technique of ref.~\sxsym. Next, we 
show how the
deep--inelastic structure functions $F_2(x,Q^2)$ and $F_L(x,Q^2)$
can be obtained by combining
resummed coefficient functions with parton distributions which satisfy
resummed evolution equations, exploiting the recent results of ref.~\ballhad.
Then, we  construct the transformation from the basis of
eigenvectors of perturbative evolution to that of quarks and gluons
in the \MS\ and \QMS\ schemes and we use it to construct the
two-by-two matrix of splitting functions in these schemes. Finally, we
discuss the phenomenological consequence of the resummation:  first we
assess the impact on  splitting functions and on
the evolution of representative quark and gluon distributions, and
then we determine the effect on the deep inelastic
structure functions $F_2(x,Q^2)$ and $F_L(x,Q^2)$.

Recently, the full  $n_f\not=0$ resummed evolution matrix has also been
constructed explicitly in ref.~\matevol, based on the approach of refs.~\refs{\ciafresb,
\salam-\ciafdip}. This result has been
obtained by extending a BFKL--like approach to coupled quark and gluon
evolution. This has the advantage of giving evolution equations for
off-shell, unintegrated parton distributions, but it has the
shortcoming of providing results in a factorization scheme which 
only coincides with \MS\ up to the
next-to-leading fixed order, and differs from it at the resummed
level: it is therefore difficult to compare our results to those of
this reference, where no predictions for physical observables are given.

\newsec{Resumming the singlet anomalous dimension matrix}

When $n_f\not=0$, we must consider
the full set of  splitting functions $P_{ij}(\as,x)$, where $i,j$ run
over quarks, gluons, or linear combinations thereof. However, the 
resummation of the full set of $P_{ij}(\as,x)$ can be obtained
from the resummation of the  largest (gluon-sector)
eigenvector of the anomalous dimension matrix
\eqn\addef{
\gamma_{ij}(\as,N) =\int^1_0\!dx\,x^{N-1}P_{ij}(\as,x),  }  
as discussed long ago
in ref.~\sxphen. The construction of the resummed anomalous dimension
when $n_f=0$ was in turn
described in detail in ref.~\sxsym.  In this section, we will
recall the general structure of the resummation of the large anomalous
dimension of ref.~\sxsym, concentrating on the aspects which require
modification when $n_f\not=0$, while referring to  ref.~\sxsym\ for
details of the resummation.

As is well known,
all nonsinglet
splitting functions
are suppressed by a power of $x$ in comparison to
the singlet~\lkns: namely, 
the rightmost singularity 
of $\gamma_{ij}$ is at $N=0$ in the singlet channel and
at $N=-1$ in the nonsinglet channel. Therefore we will henceforth only
consider the singlet sector, where the anomalous dimension is
a two-by-two matrix which provides the evolution of the Mellin moments
\eqn\mellin{f(N,t) =\int^1_0\!dx\,x^{N-1}f(x,t),}
of the parton distributions
\eqn\pdfvec{f(x,t)=\pmatrix{Q(x,t) \cr G(x,t)\cr}=\pmatrix{x q(x,t) \cr x g(x,t)\cr},}
(where $q$ and $g$ are the usual singlet quark and gluon parton densities) according to the evolution equation
\eqn\tevol{
\frac{d}{dt}f(N,t)=\gamma(\as(t),N) f(N,t), }
where $t=\ln(Q^2/\mu^2)$. 
As also well known, the reason why only one of the  eigenvalues
$\gamma^\pm(\as,N)$ 
of this matrix
needs to be resummed is that only one eigenvalue is nonvanishing at
the leading log $x$ (LL$x$) level, i.e., only one eigenvalue has a
$k$-th order pole at $N=0$ when evaluated at order $\as^k$, so 
\eqn\llxev{\eqalign{
\gamma_{{\rm LL}x}^+(\as,N)&=\gamma_s^+\left(\smallfrac{\as}{N}\right),\cr
\gamma_{{\rm LL}x}^-(\as,N)&=0.\cr}}
It follows that it is possible to choose the  factorization
scheme in such a way that
$\gamma^-(\as,N)$ is regular at $N=0$~\refs{\sxphen,\mom}. 
In particular, this is the case
 at the next-to-leading log $x$ level in the \MS\
and DIS schemes~\ch. We will henceforth only consider schemes where
$\gamma^-$ is regular at $N=0$, and thus only $\gamma^+$ has to be
resummed.

\subsec{The dominant singlet eigenvector}

The resummation of $\gamma^+$ is performed as discussed in
ref.~\sxsym. In order to understand this resummation, it is useful to
recall that the solution of the GLAP equation~\tevol\ for  the
large eigenvector $f^+$ of $\gamma$, 
\eqn\tevolpl{
\frac{d}{dt}f^+(N,t)=\gamma^+(\as(t),N) f^+(N,t), } 
coincides at leading twist with the solution to the BFKL equation
\eqn\cievol{\frac{d}{d\xi}f^+(x,M)=\chi(\ahat,M) f^+(x,M),}
where $\xi=\ln(1/x)$, $f^+(x,M)$ is the Mellin transform
\eqn\Mmom{  f^+(x,M)=\int^{\infty}_{-\infty}\! dt\, e^{-Mt} f^+(x,t),
}
$\ahat$ is the operator obtained from $\as(t)$ by the replacement
$t\to-{\partial\over\partial M}$, and the kernel\foot{Different
choices of ordering for the operator $\ahat$ in the definition of the
kernel correspond to  different choices for the argument of the
running coupling~\rcdual.}
\eqn\bfker{
\chi(\ahat,M)=\ahat \chi_0(M)+\ahat^2 \chi_1(M)+\cdots}
is determined by the kernel $\gamma$ (or conversely) by a suitable
duality relation~\refs{\afp,\sxres,\sxrun,\rcdual}. Because parton
distributions behave as a constant at large $Q^2$, while they vanish
linearly with $Q^2$ as $Q^2\to0$, the integral eq.~\Mmom\ exists when
$0<{\rm Re}M<1$ (physical region, henceforth), 
and it can be defined elsewhere by analytic
continuation.

At fixed coupling the duality relations between the kernels are simply
\eqn\dual{\eqalign{ 
\chi(\as,\gamma^+(\as,N))&=N, \cr
\gamma^+(\as,\chi(\as,M))&=M, \cr}}
which imply that if we expand 
 $\gamma^+(\as,N)$ in powers of $\as$ at fixed
$\as/N$
\eqn\sxexp
{\gamma^+(\as,N)=\gamma^+_s(\smallfrac{\as}{N})
+\as\gamma^+_{ss}(\smallfrac{\as}{N})+\cdots,} 
$\gamma^+_s$ is determined by  $\chi_0$ (and conversely), $\gamma^+_{ss}$
by $\chi_0$ and $\chi_1$ and so on. At the running coupling level it
is still true that the first $n$ orders of the expansion eq.~\sxexp\
are determined by $\chi(\as,M)$, eq.~\bfker, up to $n$-th order,
but eq.~\dual\ holds only at leading order (i.e. for $\chi_0$ and
$\gamma_s$), while  beyond the leading order it 
gets corrected by a series of terms up to
$O[(\beta_0\as)^{n-1}]$, which can be determined explicitly
order-by-order in perturbation theory~\rcdual.
Conversely, the first $n$ orders of the expansion of $\chi$ in powers
of $M$ at fixed $M^{-1}\ahat$ 
\eqn\sxcexp
{\chi(\ahat,M)=\chi_s(M^{-1}\ahat)+\ahat\chi_{ss}(M^{-1}\ahat)+\cdots} 
are determined by knowledge  up to $n$-th order 
of the expansion of $\gamma^+$ at fixed $N$ 
\eqn\gammadef{
\gamma^+(\as,N)=\alpha_s \gamma^+_0(N)+\alpha_s^2
\gamma^+_1(N)+\cdots . }
Of course, different orderings of $M^{-1}$ and $\ahat$ in the argument
of  the coefficients $\chi_{s^n}$ of the expansion eq.~\sxcexp\ lead to a different
functional form of the coefficients; the ordering eq.~\sxcexp\ is
particularly convenient because with it the leading order coefficient
$\chi_s$ has the same form as that obtained from $\gamma_0$ using
fixed-coupling duality eq.~\dual~\rcdual.


The resummation of $\gamma^+(\as,N)$ at $k$-th order consists of
supplementing the $k$-th order of its 
expansion in powers of $\as$ eq.~\gammadef\
with three further classes of terms: (a) terms up to $k$--th order in the
expansion of $\gamma^+$  in powers of $\as$ with $\as/N$ fixed, eq.~\sxexp,
(double-leading resummation, henceforth);
(b) contributions which are subleading with respect to the double-leading 
resummation but which enforce the physical
constraints of momentum conservation and gluon exchange symmetry
(symmetrization, henceforth);
(c) contributions which are subleading with respect to the double-leading 
expansion but which are needed in order to ensure the uniform
convergence of that expansion in the physical region 
(running-coupling resummation, henceforth).
Let us now discuss  the resummation of  these classes of terms in
turn.

The double--leading resummation (first step) is performed by combining
the first $k$ orders of the expansion of $\gamma^+$ in powers of $\as$
at fixed $N$ eq.~\gammadef\ and at fixed $\as/N$ eq.~\sxexp\
(double-leading expansion~\refs{\summ,\sxres}), and subtracting double
counting:
\eqn\dlgampl{\eqalign{
\gamma_{\scriptstyle\rm DL}^+(N,\as)&=\left[\as\gamma^+_{0}(N)
+\gamma^+_{s}\left(\smallfrac{\as}{N}\right)-
\smallfrac{n_c\as}{\pi N}\right]\cr
&\qquad +\as\left[\as\gamma^+_{1}(N)
+\gamma^+_{ss}\left(\smallfrac{\as}{N}\right)
-\as\left(\smallfrac{e^+_2}{N^2}+
\smallfrac{e^+_1}{N}\right)-e_0\right]+\cdots.\cr}}
The  key
observation is that one can prove~\sxres\ that the double--leading
expansion of $\gamma^+$ eq.~\dlgampl\ and  the double leading
expansion of $\chi$ are order by order dual to each other.  This is
crucial because the subsequent steps of the resummation
(symmetrization and running-coupling resummation) are performed by
manipulating $\chi$. The construction of the
resummed $\gamma^+$ thus starts~\sxsym\ by transforming the double--leading
$\gamma_{\scriptstyle\rm DL}^+(N,\as)$ eq.~\dlgampl\ into its dual
\eqn\kiDL{\eqalign{
\chi_{\scriptstyle\rm DL}&=  \left[\as \chi_0(M) +
\chi_s(\smallfrac{\as}{M})-\smallfrac{n_c\as}{\pi M}\right]\cr
&\quad+\as\left[ \as\chi_1(M)+\chi_{ss}(\smallfrac{\as}{M})-
\as\left(\smallfrac{f_2}{M^2}+\smallfrac{f_1}{M}\right)-f_0\right]+\dots.}}

The kernel $\chi$ eq.~\kiDL\ has a stable perturbative expansion for
small $M$. This is a consequence of the fact that momentum
conservation implies~\sxphen\ that $\gamma^+(1)=0$, which, by duality,
entails $\chi(0)=1$ (up to running coupling corrections). 
These properties are exactly satisfied order-by-order by the
perturbative expansion eq.~\gammadef\ of $\gamma^+$, and thus only violated
by (small) subleading terms in the double--leading expansion
eq.~\kiDL\ of $\chi$, which is thus finite (and close to one) at $M=0$
despite the fact that subsequent orders of the expansion eq.~\bfker\
of $\chi$ have poles of increasingly high order and  alternating sign
coefficients: these poles are removed by the double leading
resummation eq.~\kiDL. However, order by order in the expansion
eq.~\bfker\ the kernel $\chi$ also has poles at $M=1$. Hence, the
double--leading expansion of $\chi$ is still perturbatively unstable
when $M$ grows sufficiently large. This is problematic because 
perturbative evolution
at small $x$ is controlled by the small $N$ behaviour of $\gamma^+$,
which by duality corresponds to the large $M$ behaviour of $\chi$.

This instability can be cured by exploiting a
symmetry of the set of Feynman diagrams which determine the kernel $\chi$, that
implies that the kernel satisfies the reflection relation $\chi(M)=\chi(1-M)$ 
order by order in perturbation theory. This exchange symmetry
is
broken by an asymmetric choice of kinematic variables, as  is
necessary for the application to deep inelastic scattering,  
and by running coupling
effects~\refs{\salam,\ciafresa}. The double--leading kernel can thus be 
symmetrized by first
undoing all sources of symmetry breaking, then symmetrizing, and finally
restoring the symmetry breaking. The symmetrization must be performed
in such a way that the symmetrized kernel $\chi_\Sigma$ only differs from
the double-leading $\chi_{\scriptstyle\rm
DL}$ by subleading terms. This introduces a certain ambiguity,
which however is controlled by the requirement of momentum
conservation, which fixes the value of $\chi$ at $M=0$. The way the
symmetrization can be implemented at the leading- and next-to-leading
order of the double-leading expansion was described in detail in
ref.~\sxsym\ to which we refer, since the procedure is unchanged
regardless of the value of $n_f$. Once the resummed, symmetrized
kernel $\chi_\Sigma(M)$  has been constructed to the desired order, 
the corresponding resummed
anomalous dimension $\gamma^+_\Sigma$ can be obtained from it using
running--coupling duality, as we shall see shortly. 

An important consequence of the symmetrization of $\chi$ is that the
symmetrized kernel always has a minimum in the physical region $0\le \Re M
\le 1$, and it is an entire function
of $M$ for $\Re M>0$. The existence of the minimum is a 
generic consequence of symmetrization of the behaviour around the
momentum conservation value $\chi(0)=1$, where the kernel has  
negative derivative; this in turn is, by duality, a generic
consequence of  the physical requirement that $\gamma^+$ decreases as
$N$ increases. The fact that the kernel is an entire function then
follows from the transformation from the kinematics in
which the kernel is symmetric (appropriate for processes such as
two-jet production in $p$--$p$ scattering) to that of DIS~\sxsym. The
transformation from symmetric to DIS variables leaves unchanged the 
value of $\chi$ at the minimum  
(and the curvature of $\chi$ at the minimum~\sxsym), but shifts the 
position of the minimum away from $M=\half$ by small corrections.

The existence of a minimum of $\chi$ is what makes the third step, 
the running-coupling resummation, necessary. Indeed, the
double--leading improvement eq.~\dlgampl\ of the expansion~\gammadef\ 
of $\gamma^+$ is necessary but not sufficient for the
perturbative corrections to $\gamma^+$ to be 
uniformly small in the small $x$ limit: the inclusion of the first $k$
orders of the expansion eq.~\sxexp\ of $\gamma^+$ in powers of $\as$
at fixed $\as/N$ guarantees that the $O(k+1)$ is $O(\as)$ if $\as/N$
is kept fixed as $N$ decreases, but not if $N$ decreases at fixed
$\as$. Now, if $\chi$ has a minimum at $M=M_0$, in the vicinity of the
minimum we can expand
\eqn\finquad{\chi^q(\ahat,M)=c(\ahat)+\half
\kappa(\ahat)\left(M-M_0\right)^2+O\left[\left(M-M_0\right)^3\right].}
The fixed--coupling dual to $\chi^q$ eq.~\finquad\ is
\eqn\gamquad{\gamma_q^+(\as,N)=M_0-\sqrt{\frac{N-c(\as}{\half\kappa(\as)}},}
which has square-root branch cut at
$N=\chi^q(\as,M_0)$. However, the running coupling
correction to  $\gamma_{ss}$ in the vicinity of the minimum is
\eqn\rccorrg{\eqalign{\gamma^{\beta_0}_{ss}(\as,N)&=-\beta_0\frac{\chi_0^{\prime\prime}
(\gamma_q^+)\chi_0(\gamma_q^+)}{2[\chi_0^\prime(\gamma_q^+)]^2}\cr
&=
-\frac{1}{4}\beta_0\frac{N}{N-c(\as)},\cr}}
which has a simple pole at $N=\chi(M_0)$. 

In general, 
the running coupling correction to  $\gamma_{s^n}$ has an order $n$ pole
at $M=M_0$~\refs{\rcdual,\mf}. This implies the growth
of the associated
splitting function by a power of $\ln 1/x$ equal to the order of the
pole in comparison to a pole--free splitting function. Hence the
running coupling contributions are not uniformly small at small $x$
and they must be resummed. 

The resummation of running coupling singularities is possible~\sxrun\ thanks to
the fact that the running-coupling BFKL equation~\cievol\ can be
solved in closed form either for a quadratic kernel which is linear in
$\as$~\liprun\ or, at the leading log level, for a quadratic kernel
which is a generic function of $\as$. 
As derived in detail in refs.\refs{\runph,\sxsym}, the inverse
Mellin transforms  of these two closed form solutions can 
be written down in terms of an Airy function or a Bateman function, 
$G_A(N,t)$ or $G_B(N,t)$, respectively. It is then possible to
determine the associated anomalous dimensions as their logarithmic
derivative:  
\eqn\batandim{\gamma_B(\as,N)=\frac{\partial}{\partial t}
\ln  G_B(N,t) 
.}
The Bateman result includes the Airy case to which it reduces for a
kernel which is proportional to $\as$. 
 
The ``Bateman'' anomalous dimension eq.~\batandim\
resums to all orders the leading singular running coupling corrections
to duality. 
The leading singularity which controls the small $x$
behaviour is then the singularity of the Bateman anomalous dimension,
which is a simple pole:
\eqn\batres{\gamma_B(\as(t),N)={1\over N-N_0} {r_B}+
O[(N-N_0)^0],\qquad r_B= 2\beta_0\bar\as\left[N_0-\bar c(\as)\right],}
where $N_0$ is related to the location of the zeros of the Bateman
function in eq.~\batandim. The value of $N_0$ depends on the values of
$c(\as)$ and $\kappa(\as)$. It is plotted as a function of
$\as(t)$ at LO and NLO in ref.~\sxsym. Specifically, 
for $\as=0.2$, $N_0\approx
0.17$ at both LO and NLO. This value corresponds to a drastic
suppression of the asymptotic rise at small $x$ with respect to
the fixed coupling case. Moreover, the smallness of the associated
residue $r_B$ delays the onset of the powerlike asymptotic behaviour
to values of $x$ below the region of the HERA data. 

A resummed result which is uniformly stable at small $x$
is found by simply combining the  anomalous dimension obtained using
running coupling duality from the symmetrized kernel $\chi_\Sigma$
with the ``Bateman'' resummation eq.~\batandim\ of the running
coupling corrections, and subtracting the double counting.
Namely, at next-to-leading order, one first constructs 
\eqn\prcdual{\gamma^{+\,rc,\,pert}_{\Sigma\,NLO}(\as(t),N)=
\gamma^+_{\Sigma\,NLO}(\as(t),N)
-\beta_0\as(t)\left[\frac{\chi_0^{\prime\prime}(\gamma^+_s(\smallfrac{\as}{N}))
\chi_0(\gamma^+_s(\smallfrac{\as}{N}))}{2[\chi_0^\prime(\gamma^+_s(\smallfrac{\as}{N}))]^2}-1\right],}
where $\gamma^+_{\Sigma\,NLO}(N,\as(t))$ 
is obtained from the NLO symmetrized kernel $\chi_{\Sigma\,NLO}$ using
fixed--coupling duality eq.~\dual, and the term proportional to
$\beta_0$ is the running coupling correction.
Then the result is combined with the Bateman resummation:
\eqn\rcnlores{\eqalign{\gamma^{+\,res}_{NLO}&\equiv
\gamma^{+\,rc}_{\Sigma\,NLO}(\as(t),N)
=\gamma^{+\,rc,\,pert}_{\Sigma\,NLO}(\as(t),N)+
\gamma^B(\as(t),N)+\cr
&- \gamma^B_s(\as(t),N)-\gamma^B_{ss}(\as(t),N)+\gamma_{\rm match}(\as(t),N)
+\gamma_{\rm mom}(\as(t),N),\cr}}
where the double counting subtractions $\gamma_{B,\,s}$ and $\gamma_{B,\,ss}$ 
are defined from the expansion
\eqn\batasym{\gamma_B(\as,N)=\gamma_{B,\,s}(\smallfrac{\as}{N})+
\as\gamma_{B,\,ss}(\smallfrac{\as}{N})+O(\as^2),}
$\gamma_{\rm mom}$ is a subleading correction which enforces
exact momentum conservation, and $\gamma_{\rm match}$ is a subleading
correction which ensures that at large $N$ the resummed result exactly
coincides with the unresummed NLO result $\gamma^+_{\rm NLO}$ rather
than just reducing to it up to NNLO terms, and  specifically that all
subleading terms induced by the resummation vanish at least as $1/N$
as $N\to\infty$.

\subsec{Virtual quark effects}

The construction of the resummed anomalous dimension discussed so far
is the same as that of ref.~\sxsym, the only difference being that
when $n_f=0$ there is only one parton distribution 
and one anomalous dimension,
while  when
$n_f\not=0$ the anomalous dimension one resums is the eigenvector 
$\gamma^+$ of
the two-by-two singlet anomalous dimension matrix. 
There is a complication however. Namely, the
two eigenvalues of the leading--order 
anomalous dimension matrix $\gamma_0$ become equal
at a pair of complex conjugate values $N^d=N^d_r\pm i N^d_i$, with
$N^d_r>0$, where the square root of the discriminant of the quadratic
secular equation vanishes, \ie\ when
\eqn\discr{(\gamma_0^{qq}-\gamma_0^{gg})^2+4\gamma_0^{qg}\gamma_0^{gq}=0.}
Thus when $n_f\neq 0$, $\gamma^+_0$ has a pair of branch cuts. 
These singularities are manifestly unphysical, and indeed they
cancel in the solution of the
evolution equations.  Because they are to the right of $N=0$, 
if uncancelled they would lead to a spurious oscillation of the solution to
the evolution equations: the splitting function computed as the 
inverse Mellin transform of $\gamma_0^+$ is oscillatory at small $x$. 
After resummation, however, the cancellation
of singularities can be spoiled by subleading corrections. Even though 
these corrections are formally subleading, they can lead to 
large effects due to the small $x$ 
instabilities they induce. In order to
avoid unphysical behaviour of the solutions, we must make sure that
the resummation corrections are implemented in such a way that the 
cancellation of these singularities remains exact in the final solution, 
which in turn requires that the unphysical branch cuts in the 
anomalous dimension 
be identical to those in the fixed order anomalous dimension.

We do this in the following way. Firstly we separate off the
$n_f$ dependent resummation correction 
$\Delta^{\rm res}\gamma^{+\,res}_{NLO}$ to the
resummed anomalous dimension $\gamma^{+\,res}_{NLO}$ eq.~\rcnlores:
\eqnn\rcsep\eqnn\cut
$$\eqalignno{
\gamma^{+\,res}_{NLO}(\as,N;n_f)&\equiv\gamma^{+\,res}_{NLO}(\as,N;0)+
\gamma^{+\,cut}_{NLO}(\as,N;n_f)+ 
\Delta\gamma^{+\,res}_{NLO} (\as,N;n_f),&\rcsep\cr
\gamma^{+\,cut}_{NLO}(\as,N;n_f)&\equiv\as\gamma^+_0(N;n_f)
+\as^2\gamma^+_1(N,n_f)-(\as\gamma^+_0(N;0)+\as^2\gamma^+_1(N,0)),&\cut\cr}$$
where eq.~\rcsep\ can be viewed as a definition of 
$\Delta\gamma^{+\,res}_{NLO}$ in terms of 
$\gamma^{+\,res}_{NLO}$. All the $n_f$ dependence is now made explicit: 
$\gamma^{+\,res}_{NLO}(\as,N;0)$ is cut free, while 
$\gamma^{+\,cut}_{NLO}(\as,N;n_f)$ contains the fixed order cuts 
described above, but vanishes at $n_f=0$. The formally subleading but
asymptotically dominant 
singularities would arise if
$\Delta\gamma^{+\,res}_{NLO}(\as,N;n_f)$ were calculated using 
the $n_f\neq 0$ version of eq.~\rcnlores: the  
cut in $\gamma^{+\,cut}_{NLO}$ would propagate into $\chi_s$ and 
thus $\chi_{\rm DL}$ eq.~\kiDL. 
To circumvent this difficulty, we instead 
compute the resummation correction $\Delta\gamma^{+\,res}_{NLO}$
using a rational approximation to 
$\gamma^{+,\ cut}_{NLO}$ eq.~\cut, $\gamma^{+,\ rat}_{NLO}$, 
which is accurate when 
$N$ is small (in the resummation region), and goes away at large $N$ 
(where it is irrelevant), but is free of cuts. Thus we extend the
double--leading expansion eq.~\dlgampl\ from $n_f=0$ to $n_f\neq 0$ by 
adding to it $\gamma^{+,\ rat}_{NLO}$ instead of $\gamma^{+,\ cut}_{NLO}$, 
include the $n_f$ dependent pieces in $\chi_1$, and then follow from then 
onwards the same procedure as before as far as eqn.\rcnlores, to give us
$\gamma^{+\,res,rat}_{NLO}(\as,N;n_f)$. Subtracting the $n_f=0$ anomalous 
dimension then gives us the desired result:
\eqn\rcsepx{\Delta\gamma^{+\,res}_{NLO}(\as,N;n_f)=
\gamma^{+\,res,rat}_{NLO}(\as,N;n_f)-\gamma^{+\,res}_{NLO}(\as,N;0),}
free from cuts. This procedure ensures that the resummed anomalous dimension is
everywhere  accurate, provided only the rational approximation is
accurate in the small $N\lsim 0.5$ region where the resummation kicks in:
for larger values of $N$, where the resummed and unresummed results
coincide by construction,
and the exact $\gamma^{+}_{NLO}$ is restored, complete with cut.
 
The rational approximation itself is constructed as a systematic
improvement of the Laurent expansion of $\gamma^{+\,cut}_{NLO}$ eq.~\cut, 
about its leading singularity at $N=0$. It is based on the physical 
requirements that it must only
has singularities on the negative real axis, so the small $x$
behaviour of the associated splitting function is only modified by
terms suppressed by powers of $x$, and that it goes to a constant at
large $N$, so that no large cancellations occur when the exact large
$N$ behaviour is restored. Finally, exact momentum
conservation is imposed consistent with these requirements. In
practice, we first construct a $i$--th order approximation
\eqn\ratap{\eqalign{\bar\gamma_{(1)}^{rat}(N)&=c_{-1}\frac{1}{N}+c_0 +
c_1\frac{N}{1+N},\cr
\bar\gamma_{(2)}^{rat}(N)&=c_{-1}\frac{1}{N}+c_0 + 
c_1 \frac{N(1+2N)}{\left(1+N\right)^2}+ c_2 \frac{N^2}{\left(1+N\right)^2}, \cr
\bar\gamma_{(3)}^{rat}(N)&=c_{-1}\frac{1}{N}+c_0 +
c_1 \frac{N(1+3N+3N^2)}{\left(1+N\right)^3}+ c_2 \frac{N^2(1+3N)}
{\left(1+N\right)^3}
+ c_3 \frac{N^3}{\left(1+N\right)^3},\quad{\rm etc,}\cr}}
where the polynomials in the numerator of the rational coefficients of
$c_i$ are adjusted so that 
\eqn\ratexp{\bar\gamma_{(i)}^{rat}(N)=\sum_{k=-1}^i c_k N^k+O(N^{k+1}).}
We then impose momentum conservation while preserving eq.~\ratexp:
\eqn\momrat{\gamma_{(i)}^{rat}(N)=\bar\gamma_{(i)}^{rat}(N)-
\left(\frac{2N}{1+N}\right)\bar\gamma_{(i)}^{rat}(1).}
The rational approximations to  $\gamma^+_0(N)$  
eq.~\gammadef\ is constructed by using eq.~\momrat, with the
first $i$ coefficients $c_i$ equated to the coefficients of the
Laurent expansion of $\gamma^+_0(N)$ about $N=0$. The same procedure
can be used for $\gamma^+_1(N)$, and in fact also for the
subsequent orders $\gamma^+_k(N)$ by simply adding the necessary
negative powers of $N$ to reproduce higher order poles at $N=0$. 

Clearly, the rational approximation can be made arbitrarily accurate
at small $N$ by adding more terms; however the subsequent terms of the
expansion, as the order of the
approximation is increased, build up again the spurious cut
singularity at $N_d$. In practice, we find that the quadratic
approximation $\gamma^{2}_{rat}(N)$ is adequate: its accuracy is
better than 1\% at leading order for $N\le0.5$ and better than 8\% 
for the next--to--leading order term, which in
turn is however a few percent of $\gamma^+_{\rm NLO}$. For $N\le 0.3$ 
these figures drop to $0.2\%$ and $2\%$ respectively. Since the $n_f$ 
dependence of the resummation is itself only a small correction to the 
result when $n_f=0$, this 
is sufficient for our purposes.

Equation~\rcsep\ with the resummed anomalous dimension computed using the
rational approximation $\gamma^{+\,res}_{NLO,\,rat}$ 
now gives us a resummed  expression for the large eigenvalue of the
anomalous dimension matrix. The NLO resummation is obtained by
combining this with the standard NLO
unresummed expression for the small eigenvalue $\gamma^-$. A full
solution of the evolution equation is then obtained in terms of the
two eigenvalues, and projectors ${\cal M}_\pm(\as,N)$ on the eigenvectors of
the anomalous dimension matrix, such that
\eqn\projandim{\gamma= {\cal M}_+\gamma^++{\cal M}_-\gamma^-,
\qquad{\cal M}_++{\cal M}_-=\1,\quad {\cal M}_\pm {\cal M}_\pm={\cal M}_\pm,
\quad {\cal M}_\pm {\cal M}_\pm=0.}
The projectors eq.~\projandim\ have the form~\sxphen
\eqn\projexp{\eqalign{{\cal M}_+&={1\over\gamma^+-\gamma^-}
\left(\matrix{\gamma_{qq}-\gamma^-&\gamma_{qg}\cr
X &\gamma^+-\gamma_{qq}\cr}\right),\cr
{\cal M}_-&={1\over\gamma^+-\gamma^-}
\left(\matrix{\gamma^+-\gamma_{qq}&-\gamma_{qg}\cr
-X &\gamma_{qq}-\gamma^-\cr}\right),}}
where $X=(\gamma^+-\gamma_{qq})(\gamma_{qq}-\gamma^-)/\gamma_{qg}$.

\subsec{Resummed eigenvalues}

In summary, we have completely specified the resummed form of the
eigenvalues of  
the anomalous dimension. The small 
eigenvalue $\gamma^-$ is not modified by the 
resummation and thus it is determined by fixed order perturbation
theory. For the  
large eigenvalue the resummed result is obtained by combining eq.~\rcnlores\ 
with eqs.~\rcsep. All quantities appearing in these equations are 
either defined directly in this paper or in ref.~\sxsym. From the expressions 
of the eigenvalues of the anomalous dimension we can construct the solutions 
of the evolution equations by using the projector formalism described above.
The projectors ${\cal M}_\pm(\as,N)$ are fully determined
from the knowledge  
of  the eigenvectors $\gamma^\pm$ and the quark--sector matrix
elements $\gamma_{qq}$ and $\gamma_{qg}$. 

The resummation
of the quark sector anomalous dimensions requires understanding, at
the resummed  
level, firstly 
how to combine coefficient functions with
evolved parton distributions, and secondly how to select the
factorization scheme. We will discuss both issues  in the
following two sections.

\newsec{Resummation of physical observables}

Resummed physical observables are obtained by combining parton distributions
which obey resummed evolution equations with resummed coefficient
functions. In the specific case of deep--inelastic scattering, we
consider the flavour singlet component of the 
structure functions $F_2(x,t)$ and $F_L(x,t)$. Because, as already
mentioned, in the nonsinglet channel the leading singularity at small
$x$ is suppressed by a power of $x$, we will not consider nonsinglet
contributions and henceforth we will denote with $F_i$ the singlet part
of the structure functions. The 
$N$--Mellin transform of the structure functions
\eqn\strfmel{F_i(N,t)=\int^1_0\!dx\,x^{N-1}\!F_i(x,t),}
can generally be written in $N$ space as
\eqn\strfundef{\eqalign{F_i(N,t)&=c^i_q(N,\as(t))
Q(N,t)+c^i_g(N,\as(t)) G(N,t)\cr &=c(N,\as(t)) f(N,t),}
}
where $f(N,t)$ is the parton distribution eq.~\mellin,\pdfvec, and
the elements $c^i_j(N,\as(t))$ ($i=2,l$, $j=q,g$) of the matrix $c(N,\as(t))$ of
 coefficients  are moments of a
partonic cross section  
\eqn\cfundef{c^i_j(N,\as(t)) =\int^1_0\!dx\,x^{N-1}\,c^i_j(x,\as(t)).}

\subsec{Double leading expansion of the coefficient functions}

In order to construct the small $x$ expression of the coefficient functions we 
exploit the fact that the factorized expression eq.~\strfundef\ can be derived 
from a more general
 $k_T$--factorization (or high--energy factorization) formula~\hefac.
 This factorization formula reduces to the more familiar $k_T$ integrated form 
in the collinear limit, and is useful for our present purposes. The 
$k_T$--factorized form of structure functions is conveniently written 
in terms of the double Mellin
 transform eqs.~\mellin,\Mmom\ 
of $k^2$--dependent coefficient functions $\cal C$ and parton
distributions  $\cal F$:
\eqn\ktfaceq{F_i(N,M)={\cal C}^i_q(N,M,\ahat) {\cal Q}(N,M;\mu^2)+{\cal
 C}^i_g(N,M,\ahat)  
{\cal G}(N,M;\mu^2),}
i.e., in vector notation:
\eqn\mateq{F(N,M)={\cal C}(N,M,\ahat){\cal F}(N,M;\mu^2).}
The $k^2$--dependent coefficient functions are
\eqn\cmel{{\cal C}^i_j(N,M) =\int \frac{d k^2}{k^2}
 \left(\frac{Q^2}{k^2}\right)^{-M}
\sigma^i_j\left(\frac{Q^2}{k^2},N\right),}
the Mellin transform of the partonic $i$-th structure function
 ($i=2,\,L$) for scattering
 of a virtual photon with virtuality $-Q^2$ off a parton $j$ ($j=q,\,g$)
 with virtuality $-k^2$.\foot{The
 factorization formula of ref.~\hefac\ is yet more general, in
 that  it applies to  coefficient functions and parton distribution 
which depend on $\vec
 k$. Here we are not interested in the angular dependence and thus we
 discuss only the more restrictive version which is obtained from it
 after averaging over the angular dependence of $\vec k$ and only
 retaining the dependence on $k^2$.} 
  
The coefficient function eq.~\cfundef\ is obtained from eq.~\ktfaceq\ 
using collinear factorization, i.e. extracting the collinear
 contribution to the  Mellin inversion integral
\eqn\cfunim{F_i(N,t)= \int_{c-i\infty}^{c+i\infty} \frac{dM}{2\pi i}
e^{M t}  F_i(N,M),}
where the integration path runs to the right of the singularities near 
$M=0$ in
the complex $M$ plane.
After factorizing all collinear logs in the evolution of the parton
distribution, the coefficient function has a single collinear log,
related to the integration over the transverse momentum of the
corresponding off-shell parton line. Because upon $M$-Mellin transform
powers of $t=\ln\frac{Q^2}{\mu^2}$
 become powers of $\frac{1}{M}$, this means that the $(M,N)$ space
$k_T$--factorized   parton
distributions $\cal{F}(N,M)$ 
have multiple $M=0$ poles, whereas the coefficient functions
${\cal C}^i_j(N,M)$ have a simple pole. The contribution of these poles to
the Mellin inversion integral
eq.~\cfunim\ then gives the collinear factorized result
eq.~\strfundef, with\eqnn\collcffromkt\eqnn\smallcdef
\eqn\smallcdef{c^i_q(N,\as(t)) = C^i_j(N,M)\Big|_{M=0},\qquad
C^i_j(N,M)\equiv M{\cal C}^i_j(N,M),}
and
\eqn\pdffac{ f(N,t)=\Gamma(N,t)f(N,0),}
where $\Gamma(N,t)$ contains the contribution of the
collinear $M$ poles. The
initial parton distribution, evaluated at $Q^2=\mu^2$ (i.e. $t=0$), is given by 
\eqn\intunint{f(N,0)={\cal F}(N,M;\mu^2)\Big|_{M=0},}
as follows from an evaluation of the integral eq.~\cfunim\ at $Q^2=\mu^2$,
where the parton distribution is free of poles.

 The parton distribution
satisfies the evolution equation~\tevol\ thanks to the $t$ dependence
due to the $M=0$ poles included in the factor $\Gamma(N,t)$:
specifically, eq.~\tevolpl\ for the large eigenvector~\llxev\ implies
\eqn\mnglap{\eqalign{\frac{d}{dt} 
\Gamma^+(N,t)f^+(N,\mu^2) 
&= \int_{c-i\infty}^{c+i\infty}\frac{dM}{2\pi i}
e^{M t} M  f^+(N,M)
\cr
&=\gamma^+(\ahat,N) \int_{c-i\infty}^{c+i\infty}\frac{dM}{2\pi i}
e^{M t}f^+(N,M).}}
It is useful to view this as an equation satisfied by 
$f^+(N,M)$ at the collinear pole~\rcdual:
\eqn\opglap{ Mf^+(N,M)=\gamma^+(\ahat,N)f^+(N,M).}
The GLAP evolution equation~\opglap\ is, of course, satisfied to any
logarithmic order thanks to the fact~\cfp\ that 
collinear singularities can be factored into the parton distribution.
Hence, $\gamma^+(\ahat,N)$ in eq.~\opglap\ is an operator which at the leading
(next-to-leading, \dots) $\ln Q^2$ level can be determined trivially
from the GLAP kernel. Furthermore, using running--coupling
duality~\rcdual, $\gamma^+(\ahat,N)$ can also be determined
at the leading
(next-to-leading, \dots) $\ln x$  level from the
BFKL kernel.

High--energy factorization thus reproduces collinear
factorization,  
in that the coefficient function thus obtained 
coincides with the collinear coefficient function eq.~\collcffromkt, while the parton
distribution satisfies the GLAP equation~\opglap. 

However, here the main interest of $k_T$ factorization is that at high energy
eq.~\ktfaceq\ holds
even away from the collinear $M=0$ pole, namely, the Mellin inverse 
eq.~\cfunim\ of the  $k_T$-factorized structure function eq.~\ktfaceq\ gives the
correct expression for the structure function, at least at the leading
$\ln x$ level. This is sufficient for the determination~\ch\
of the structure functions to the next-to-leading~$\ln x$ level, and
thus for the construction of the next-to-leading order of the
double--leading expansion~\sxphen\ of the coefficient functions, 
because 
the coefficients ${\cal C}^i_j(N,M)$ have the form
\eqn\cfsxexp{{\cal C}(N,M)= \frac{1}{M}\left[ C_{0} + \ahat
C_{1}(N,M)+O(\ahat^2)\right],}
with $C_{0}={1\,0\choose0\,0}$.

The expansion of $c(N,\as(t))$ in powers of $\as(t)$ at fixed
$\as(t)/N$ can then be determined to first nontrivial order (i.e. to
next-to-leading~$\ln x$ order) using eq.~\opglap.
Indeed, by dividing and multiplying by $M$ the right--hand side of the factorized expression~\mateq\  and performing the Mellin inversion integral, we get
\eqn\ktfactf{F(N,t)= C_0 f(N,t)+  \as(t)
\int_{c-i\infty}^{c+i\infty}\frac{dM}{2\pi i}e^{M t} 
C_1(N,M) f(N,M)+O(\as^2(t)),} 
where $f(N,t)$ is the parton distribution eq.~\pdffac, $f(N,M)$ its Mellin
transform~\Mmom, and 
$C_0$, $C_1$ are as in eq.~\cfsxexp. 

However, the GLAP equation~\tevolpl,\opglap\ implies that, at the
collinear pole, the integrals
\eqn\sqdef{ I_k\equiv \int_{c-i\infty}^{c+i\infty}\frac{dM}{2\pi i}e^{M t} 
M^k f(N,M)} 
are equal to:
\eqn\iterglap{\eqalign{I_1&=\gamma(\as(t),N)f(N,t),
 \cr
I_2&=\left(\gamma^2+\dot\gamma\right)f(N,t),
 \cr
I_3 &= \left(\gamma^3+3\gamma\dot\gamma+\ddot \gamma\right)f(N,t), 
\qquad{\rm etc.}\cr}}
The dot denotes differentiation with respect to $t$ and $\gamma$ is the 
anomalous dimension matrix acting on the parton distributions $f(N,t)$
eq.~\pdffac. Of course, all
contributions proportional to $t$ derivatives of $\gamma$ on the
right--hand side of eq.~\iterglap\ vanish at the fixed coupling level.
Note that eq.~\ktfactf\ only holds at the leading $\ln x$ level, but
eqs.~\sqdef-\iterglap\ follow from the GLAP equation and thus hold at
all logarithmic orders.

If $\gamma(\as,N)
=\gamma_s\left(\frac{\as}{N}\right)$ (leading~$\ln x$), then $\dot\gamma$ is 
next-to-leading~$\ln x$ and so forth: hence all terms beyond the first
in round brackets on the r.h.s. of eq.~\iterglap\ are subleading, so
up to subleading terms
$I_n=\gamma^n f$. Furthermore, $C(N,M)$ is regular as $N\to0$ because
all $N$--poles i.e. all logs of $x$ are included in the
evolution: we can thus set $N=0$ in it, since positive powers of $N$
lead to subleading contributions. It
follows that  the expansion of $c(N,\as(t))$ in powers of $\as$ at
fixed $\as/N$ is
\eqn\xnllx{c(N,\as(t))=C_0+
\as(t)
c_{ss}\left(\gamma^+_s\left(\smallfrac{\as(t)}{N}\right)\right){\cal M}_+
+O(\as^2)} 
where ${\cal M}_+$ is the projector eq.~\projandim, and 
\eqn\conedef{c_{ss}(M) = C_1(0,M).} In order to derive eq.~\xnllx\ we 
inserted $\1={\cal M}_++{\cal M}_-$ in eq.~\ktfactf\ between $C_1$ and 
$f$  and recalled that $\gamma^-_s=0$ (see eq.~\llxev).
The explicit expression for  
the four matrix elements $c_{ss}(M)$ in the
\MS\ scheme is given in
ref.~\ch. 

The double leading expression for the coefficient function
  is obtained combining the first 
$k$ orders of the expansion of $c(N,\as(t))$ in powers of $\as$
 at fixed $\as/N$ with the first $k$ orders of the standard expansion
  of $c(N,\as(t))$ in powers of $\as$ at fixed $N$, and subtracting double
counting: up to next-to-leading order
\eqn\dlcf{c_{\rm DL}(N,\as(t))= C_0
+\as(t)\left[(c_1(N)-c_1(0))+
c_{ss}\left(\gamma^{+\,res}_{\rm NLO}\left(\as(t),N\right)\right)
{\cal M}_+\right] + \cdots,}
where $c_1(0)$ is the subtraction for double counting, 
and $\gamma^{+\,res}_{\rm NLO}$ is the resummed anomalous dimension
  eq.~\rcsep. The identification of the second argument of  $c_{ss}$
with $ \gamma^{+\,res}_{\rm NLO}$ follows from
eq.~\iterglap\ with $\gamma=\gamma^{+\,res}_{\rm NLO}$,  neglecting all
 subleading  terms with $t$ derivatives of $\gamma$. Of course,
the 
 replacement of the leading~$\ln x$ expression 
$\gamma^+_s$ with  
$ \gamma^{+\,res}_{\rm NLO}$ is also subleading.   Nevertheless, this
  replacement is needed in order to ensure that the coefficient
  function in $N$ space doesn't develop a spurious singularity at the
  location of the cut in $\gamma^+_s$, which, being to the right of
  the singularity (simple pole) which dominates the small $x$
  behaviour of $ \gamma^{+\,res}_{\rm NLO}$ (see fig.~6 of ref.~\sxsym),
  would lead to large, spurious small $x$ corrections.

\subsec{Running coupling effects}

We have seen in Sect.~2 that 
running coupling corrections to the anomalous dimension
start at next-to-leading~$\ln x$~\rccorrg, but their all--order
resummation is necessary to obtain a stable small $x$ limit,   because
their leading singularities are  of increasingly higher order at
higher orders in $\beta_0 \as$.
The all--order 
resummation of these singular terms actually changes the nature of the
leading singularity which
 dominates the small $x$ behaviour of the anomalous dimension and
associated splitting function. 

Running coupling corrections to the coefficient
functions,  namely the terms proportional to $t$
derivatives of the anomalous dimension in eq.~\iterglap, 
start at next-to-next-to-leading~$\ln x$. It is easy to see that they
also have leading singularities of increasingly high order.
Indeed, differentiating the duality relation~\dual\ with respect to
$t$ we get
\eqn\rccsing{\dot\gamma^+(\as,N)=-\frac{\dot
\chi(\as,M)}{\chi^\prime(\as,M)}\bigg|_{M=\gamma(\as,N)}, }
where the dot denotes differentiation with respect to $t$
and the prime indicates differentiation with respect to $M$. This expression is
clearly singular at the minimum of $\chi$: indeed, after running--coupling
resummation of the anomalous dimension, $\gamma$ has a simple pole
at $N_0(\as(t))$ eq.~\batres. Hence, subsequent $t$ derivatives of
$\gamma$
will have higher order poles there. It follows that the
double--leading result eq.~\dlcf\ is not stable at small $x$: the
running--coupling corrections to the coefficient function must also be
resummed to all orders. This can be done using the technique developed
in ref.~\ballhad, and (as in the case of the anomalous
dimension) 
it
changes the nature if the singularity which dominates the small $x$
behaviour of the coefficient function.

After factorization of the collinear $M=0$ poles, the 
coefficient functions $c_{ss}(M)$ have simple poles on the
real $M$ axis for positive and negative values of $M$~\ch. 
Those for negative values of $M$ correspond to
higher-twist corrections and are of no concern. Those for positive
values of $M$ start at $M=1$, i.e. at the edge of the physical region
for the Mellin inversion integral eq.~\Mmom \ and thus also seem
immaterial. Nevertheless,  if we let
$M=\gamma^{+\,res}_{\rm NLO}$ as in eq.~\dlcf, then at small $x$ the
resummed anomalous dimension is dominated by the pole eq.~\batres, so,
using the form eq.~\batres\ of $\gamma(\as,N)$, 
a simple pole in the coefficient function at $M=k$ becomes
\eqn\simpol{\frac{1}{k-M}\bigg|_{M=\gamma(\as,N)}
=\frac{(N-N_0)/k}{N-(N_0+r/k)},
}
which has a pole to the right of the pole of the anomalous
dimension. The pole in the coefficient function at $M=1$, after running
coupling resummation of the anomalous dimension but in the absence of
running coupling resummation of the coefficient function, would thus
become the leading small $x$ singularity. This is due to the fact that
the identification $M=\gamma^+(\as,N)$ in the coefficient function 
is equivalent to dominating the Mellin inversion integral eq.~\ktfactf\
with a simple pole at $M=\gamma^+(\as,N)$, but if $\gamma^+$ has the form
eq.~\batres, as $N$ decreases this pole actually moves to the right of
$M=1$ thereby pinching the path of $M$ integration.  Note that at the fixed
coupling level this does not happen because the anomalous
dimension at small $x$ then has the form eq.~\gamquad, which is
bounded by $M_0$ and thus does not lead to any new
singularities. 

However, the singularity eq.~\simpol\ is  entirely
spurious: indeed, it can be shown~\ballhad\ that the
Mellin inversion integral for a coefficient function which has a
simple pole in $M$ and a parton distribution which satisfies the
running--coupling GLAP equation~\mnglap\ has no new $N$
singularities on top of those already present in the parton
distributions. 
In particular, we can explicitly check this statement by performing
the integral in the case of a coefficient function given by a
simple pole and an anomalous dimension linear in $\as$. In fact,
taking $\gamma^+=\as(t)\gamma_0^+$, and using the 
leading order $\beta$ function $\beta(\as)=-\beta_0\as^2$, the
integral for any 
parton density $f$ can be computed exactly~\ballhad:
\eqn\exrccf{J(N,t)\equiv\int_{c-i\infty}^{c+i\infty}\frac{dM}{2\pi i}e^{M t}
\frac{1}{1-M} f(N,M)= t^{-\gamma_0(N)/\beta_0} e^t
\Gamma(1+\gamma_0(N)/\beta_0,t)f(N,t),}
where $\Gamma(x,t)$ is the incomplete Gamma function. 
Manifestly, $J(N,t)$ eq.~\exrccf\ is  free of $N$ singularities
besides those already present in $f(N,t)$. 
In this case, 
it can further be shown explicitly that the series of running
coupling contributions on the right--hand side of eq.~\iterglap\ is an
asymptotic expansion of the exact singularity--free result. Indeed, in
this case higher order $t$  derivatives of $\gamma$ can all be
expressed in terms of $\dot\gamma$ and $\gamma$, and
eq.~\iterglap\ becomes
\eqn\squbr{I_k=[\gamma^k]f(N,t),}
where we have defined $[\gamma^k]$ as follows:
\eqn\iterglaplo{\eqalign{[\gamma]&=\gamma(\as(t),N), \cr
[\gamma^2]&=\gamma^2 \left(1+\frac{\dot\gamma}{\gamma^2}\right),   
 \cr
[\gamma^3]&=\gamma^3 \left(1+\frac{\dot\gamma}{\gamma^2}\right)
\left(1+2\frac{\dot\gamma}{\gamma^2}\right),\qquad{\rm etc,}\cr}}
so that recursively
\eqn\iterrec{[\gamma^n] = 
\gamma \left(1+(n-1)\frac{\dot\gamma}{\gamma^2}\right) [\gamma^{n-1}].}
It is then easy to see that if we determine $J(N,t)$ eq.~\exrccf\
by integrating the series expansion of $\frac{1}{1-M}$ term by term
\eqn\exprccf{J(N,t)=\sum_{k=0}^\infty [ \gamma^k]f(N,t)}
and use eq.~\iterglaplo\ we recover eq.~\exrccf\ with
the standard asymptotic
expansion of $\Gamma(x,t)$ in inverse powers of $t$.

In summary  the all-order resummation of the increasingly more
divergent running coupling corrections eq.~\iterglap\ removes the spurious
singularity eq.~\simpol\ which is produced if these  running
coupling terms are neglected but the running coupling resummation is
included in the anomalous dimension. 
We could thus proceed analogously to 
the running coupling resummation of the anomalous dimension:
subtract the $M$ poles from the coefficient function, resum them using
the exact result of ref.~\ballhad, and treat the remaining regular
part of the coefficient function in the $M=\gamma$  approximation, the
corrections to which are now genuinely subleading. However, in the
\MS\ scheme, only the $F_L$ coefficient functions, i.e. the second
column of the matrix  $c_1(M)$ are known in closed form, whereas for
the $F_2$ coefficient functions  only a
series expansion in powers of $M$ is known~\ch, and this procedure
is not viable.

However, we may instead  use the explicit result
eq.~\iterglaplo\ to perform the inverse Mellin
transform order by order using the power series expansion of the
coefficient function. Indeed, eq.~\iterglaplo\ was derived assuming the
leading order form of $\gamma$ and the running of $\as$. Hence, for a
generic anomalous dimension $\gamma(N,t)$ it holds up to $O(\as^2)$
corrections.  Furthermore, we know that, for a coefficient function
which has simple poles, if running coupling terms are
included to all orders in $\as$ 
(in the expanded case using eq.~\iterglap) the coefficient function
does not have any singularities. Because the result to $O(\as)$ found
using eq.~\iterglaplo\ is already free of singularities, it follows
that the remaining corrections are genuinely subleading. 
We conclude that a running coupling resummation of $c_{\rm DL}$
eq.~\dlcf\ is given by
\eqn\rescf{\eqalign{c_{\rm DL}(N,\as(t))= C_0
&+\as(t)\left\{(c_1(N)-c_1(0))+\left[
c_{ss}\left([\gamma^{+\,res}_{\rm NLO}]\right)\right.\right.\cr
&\left.\left.\quad -c_{ss}([\gamma^{+}_{0}(N)+\as
\gamma^{+}_{1}(N)-\smallfrac{\as}{N}\left(\smallfrac{n_c}{\pi}- \as
e_1^+\right)-e_2^+\smallfrac{\as^2}{N^2}])
\right]{\cal M}_+\right\} ,\cr}}
where by $c_{ss}([\gamma^{+\,res}_{\rm NLO}])$ we mean that $c_{ss}(M)$ is expanded in
powers of $M$, and then evaluated by replacing $M^k$ with $[\gamma^k]$ 
and using eq.~\iterglaplo\ with $\gamma=\gamma^{+\,res}_{\rm NLO}$. 
The term on the last line is a matching term: it has been 
chosen to ensure that at large $N$, where $\gamma^{+\,res}_{\rm NLO}$
reduces to $\gamma^+_{0}(N)+\as\gamma^+_{1}(N)$,
the coefficient reduces to 
its standard large $N$ form, while at small $N$ it reduces to
$c_{ss}([\gamma^{+\,res}_{\rm NLO}])$, as it should.

In practice, use of eq.~\rescf\ is subject to the limitation that the
series expansion that one obtains is only asymptotic, and it must thus
be truncated after a finite number of terms. Clearly, the divergence
sets in earlier at small $Q^2$ and small $x$, where the anomalous
dimension $\gamma(N,\as)$ grows large, Unfortunately, it turns
out that for typical values of the kinematic variables, such as
$x\lsim10^{-3}$, $\as\gsim 0.2$, the series for the matrix elements 
already starts diverging
after a rather small number of terms, which leads to an
unacceptably large ambiguity in the results. 

However, we can treat the
asymptotic series by Borel resummation. Namely, the $n$-th order
contribution to eq.~\iterglaplo\ is written as
\eqn\interglbor{[\gamma^n]=\int_0^\infty \!ds\,  K(s)  \frac{s^n}{f(n)} 
\gamma^n
\left(1+\frac{\dot\gamma}{\gamma^2}\right)\dots\left(1+(n-1)\frac{\dot\gamma}{\gamma^2}\right),}
where $K(s)$ is chosen in such a way that $f(n)$ grows with $n$,
thereby improving the convergence of the series. For instance, the
standard choice is
\eqn\borelst{ K(s)=e^{-s},\quad f(n)=n!.}
With this choice,   
 in the specific case of $J(N,t)$ eq.~\exrccf\ the divergent series
eq.~\exrccf\ becomes convergent, and  one gets
\eqnn\borelcflo\eqnn\borelcdef\eqnn\borelcform
$$\eqalignno{J(N,t)&=c(N,t) f(N,t),&\borelcflo\cr
c(N,t)&=\int_0^\infty \!ds\, e^{-s} c(N,t;s),&\borelcdef\cr
c(N,t;s)&=
\left[1-s \frac{\dot\gamma(N,\as(t))}{\gamma(N,\as(t))}
\right]^{-\gamma^2/\dot\gamma}.&\borelcform \cr}$$

However, the expansion of $c(N,t;s)$ eq.~\borelcflo\ in powers of $s$
has finite radius of convergence. Because in general we cannot sum the
series in closed form, this would prevent a numerical evaluation of
the integral over $s$ eq.~\interglbor. We can solve this problem by
choosing $K(s)$ in such a way that $f(n)$ grows more than factorially
with $n$. Specifically, we 
make the choice
\eqn\dblborelcdef{ K(s)=K_0(\sqrt{s});\quad f(n)=(n!)^2,}
where $K_0$ is a modified Bessel function. Now
\eqn\dblborelcflo{c(N,t)=\int_0^\infty \!ds\,K_0(\sqrt{s})  \bar c(N,t;s),}
and the expansion of $\bar c(N,t;s)$ has infinite radius of
convergence, because its $n$--th order term is factorially smaller
than the corresponding term in the series expansion of $c(N,t;s)$
which had finite radius of convergence.
In the particular case of $J(N,t)$ eq.~\exrccf, the function $\bar c(N,t;s)$
eq.~\dblborelcflo\ is given by
\eqn\dblborelcform{
\bar c (N,t;s)=M\left(\smallfrac{\gamma^2}{\dot\gamma},1,s
\smallfrac{\dot\gamma}{\gamma}\right),}
where $M(x,y,z)$ is the confluent hypergeometric function.

We can thus use eq.~\dblborelcflo\ to evaluate the series expansion of the
coefficient function eq.~\rescf, at
 least to the extent that the dominant singularities of the
 coefficient function are poles.
In practice, since we only know the expansion of $c_{ss}$ in powers of
 $M$, we cannot sum the power series expansion in closed form before
 performing the Borel inversion integral over $s$ eq.~\borelcflo. This seems
 problematic because the
 integral of any finite--order truncation of the expansion of the
 right--hand side of eq.~\borelcflo\ leads back to the original
 divergent series. Nevertheless, if the Borel inversion integral
 exists, then one can obtain an arbitrarily accurate approximation to
 it by integrating up to a given cutoff $\Lambda$. But because the
 series has infinite radius of convergence, for any finite value of
 $\Lambda$ the integrand can be determined to any desired accuracy by
 including a finite number of terms in the series. In practice,
 instead of
 truncating the $s$ integration at some large cutoff value, we may 
 replace $\bar c(N,t;s)$ with  an extrapolation for $s>\Lambda$,
 either linear, or based on the hypergeometric form eq.~\dblborelcform.

\subsec{Resummed coefficient functions}
Summarizing, we constructed a resummed coefficient function based on
the running--coupling resummation of the double leading coefficient
function eq.~\dlcf.  
  The result has the form eq.~\rescf, where the running coupling
resummation is effected by evaluating the argument of the next-to-leading
$\ln x$ coefficient functions $c_{ss}$  by means of eq.~\iterglaplo.
The ensuing divergent series are summed through Borel transformation
eq.~\dblborelcflo. The Borel transformed series is truncated and summed
numerically, and the Borel inversion  integral is also performed
numerically by integrating up to a cutoff $\Lambda$, whence it is
extrapolated using the form of eq.~\dblborelcform, which gives the
contribution to the series from the dominant $M=1$ pole. 
The accuracy of the result can is verified by checking
 its approximate independence of the chosen value of the cutoff
$\Lambda$.

\newsec{Factorization schemes and the quark sector}

A factorization scheme change is a  redefinition of the parton
distribution of the form
\eqn\fsc{f^\prime(N,t)=U(N,\as(t))f(N,t).}
Invariance of physical observables them implies that coefficient functions
transform as
\eqn\csc{C'(N,t)=C(N,t)U^{-1}(N,\as(t)).}
The anomalous dimensions in the primed and unprimed schemes eq.~\fsc\
are then related by
\eqn\gsc{\gamma^\prime(N,\as(t))=
U(N,\as(t))\gamma(N,\as(t)) U^{-1}(N,\as(t))
+\frac{d}{dt}U(N,\as(t))U^{-1}(N,\as(t)).}

\subsec{Resummed factorization schemes}

The theory of factorization scheme changes at the resummed level
has been developed in
ref.~\mom, and in ref.~\sxphen\ at the double leading level; we recall
some of the main results. If we consider the effect of
a change of factorization scheme 
both on the expansion of anomalous dimension in powers of $\as$ at
fixed $N$ eq.~\gammadef\ and at fixed $\as/N$ eq.~\sxexp, the
requirement that a change in factorization scheme leaves the
leading--order anomalous dimension invariant allows a wider class of
factorization  scheme changes at small $x$. This is a consequence of
the fact that the leading~$\ln x$ anomalous dimension matrix has the
form
\eqn\singstruc{
\gamma_{s}=\pmatrix{0&0\cr\gamma_s^{gq}\left({\alpha_s\over
N}\right)&\gamma_s^{gg}\left({\alpha_s\over N}\right) \cr},}
where the nonvanishing entries satisfy the colour-charge relation
\eqn\gcolch{\gamma_s^{gq}= \smallfrac{C_F}{C_A}\gamma_s^{gg}.} 
It follows that if parton distributions are redefined by a
leading~$\ln x$ function $U_s(\smallfrac{\as}{N})$,
the leading~$\ln x$ anomalous dimension eq.~\singstruc\ is unchanged
provided only~\mom
\eqn\llxsch{
U_s\left(\smallfrac{\as}{N}\right)=\pmatrix{1&0\cr 
\smallfrac{C_F}{C_A} z^{gg}_s(\smallfrac{\as}{N}) &
z^{gg}_s(\smallfrac{\as}{N}) \cr},}
where $z^{gg}_s(\smallfrac{\as}{N})=1+z^{gg}_{s,\,1} \smallfrac{\as}{N}+\dots$~.

The scheme change eq.~\llxsch\ is thus a redefinition of the small $x$
normalization of the gluon distribution, and it changes  $\gamma^{qg}$
and $\gamma^+$  according to 
\eqn\schchgam{\eqalign{
\gamma^{qg\,\prime}_{ss}&=\gamma^{qg}_{ss}/u(\gamma_s^+)\cr
\gamma^{+\,\prime}_{ss}&
=\gamma^{+}_{ss}+{\beta_0\over 4\pi} 
{\chi_0(\gamma_s^+)\over \chi_0^\prime(\gamma_s^+)}
\frac{d\ln u}{dM}\Big\vert_{M=\gamma_s^+},\cr
}}
where the function $u(M)$ is defined by the implicit equation
\eqn\udef{z^{gg}_s(\smallfrac{\as}{N})=u(\gamma_s(\smallfrac{\alpha_s}{N})).}

Further small $x$ scheme changes have the standard next-to-leading
form
\eqn\dlsch{U(N,\as)=\1+\as  z_{ss}\left(\smallfrac{\as}{N}\right),}
where $z_{ss}(0)=0.$
The scheme change eq.~\dlsch\  
affects the next-to-leading~$\ln x$ anomalous dimension
according to
\eqn\sxscheme{\gamma^\prime_{ss}=\gamma_{ss}+[z_{ss},\gamma_s].}
The form eq.~\singstruc,\gcolch\ of $\gamma_s$
implies that the effect on the anomalous dimensions of
a scheme change of the form of eq.~\dlsch\
is particularly simple~\mom:
\eqn\llxsch{\gamma^\prime=\gamma+\pmatrix{\frac{C_F}{C_A}
z_{ss}^{qg}\left(\frac{\as}{N}\right) & z_{ss}^{qg}\left(\frac{\as}{N}\right) \cr
\smallfrac{C_F}{C_A}\left(z_{ss}^{gq}\left(\frac{\as}{N}\right)-z_{ss}^{qg}
\left(\frac{\as}{N}\right) \right)-z_{ss}^{gg}\left(\frac{\as}{N}\right)
 & \smallfrac{C_F}{C_A}
z_{ss}^{qg}\left(\frac{\as}{N}\right)
 \cr}\gamma_s^{gg}.}
This is useful since the component $\gamma_{gq}$ of
$\gamma_{ss}$ in fact has no effect on a calculation at the
next-to-leading order of the resummed or double--leading expansion~\sxphen, so 
in practice at NLO it is only necessary to fix $z_{ss}^{qg}$ .

It is interesting to observe that the form 
eq.~\singstruc,\gcolch\ of $\gamma_s$,
together with the requirement eq.~\llxev\ that only one of the two
eigenvalues of $\gamma$ has leading small $x$ singularities, also entails
some scheme--independent relations between
singular contributions to matrix elements of $\gamma$.
Indeed, the trace and determinant conditions for
the eigenvalues $\gamma^\pm$, if the singular contributions to
$\gamma^-$ vanish, imply
\eqn\trdet{ \gamma^{gg}+\gamma^{qq}=\gamma^+;\quad
\gamma^{gg}\gamma^{qq}=\gamma^{qg}\gamma^{gq}. }
These relations clarify the meaning of the colour-charge relation.
Indeed, to next-to-leading~$\ln x$ the second of eq.~\trdet, when combined 
with eq.\gcolch, implies that
\eqn\colch{
\gamma_{ss}^{qq}(\smallfrac{\as}{N})= 
\smallfrac{C_F}{C_A}(\gamma_{ss}^{qg}(\smallfrac{\as}{N})-\gamma_{ss}^{qg}(0)).}
To higher  orders, the colour charge relation is sufficient but not
necessary for eq.~\trdet\ to hold.

\subsec{Construction of the \MS\ and \QMS\ schemes.}

Let us now turn to the explicit construction of factorizations
schemes,
in order to construct resummed physical observables.
To this purpose, we first 
summarize the available information on the coefficient
functions and quark--sector anomalous dimensions in various
factorization schemes. 
The leading small $x$ coefficient function matrix (i.e. the function
$c_{ss}(M)$ eq.~\conedef), as well as the anomalous dimension $\gamma_{qg}$
(and thus $\gamma_{qq}$, by eq.~\colch)
have been determined at leading order 
in ref.~\ch\ in both the \MS\ and DIS schemes.
In particular, $\gamma_{qg}$ takes the form
\eqn\hdef{\gamma^{qg}_{ss}(\smallfrac{\as}{N})
= h\left(\gamma^+(\smallfrac{\as}{N})\right),}
where the function $h(M)$ may be extracted from the computation of the
quark--antiquark production cross section from photon--virtual gluon
fusion, using the high--energy factorization techniques discussed in sect.~3.2.
In the DIS scheme the function $h(M)$
is known in closed form, while 
in \MS\ $h(M)$ and $c_{ss}(M)$ are known to all orders in 
their power series expansion about $M=0$.

The DIS scheme is defined by the
identification of the structure function $F_2$ with the quark distribution:
$F_2(x,Q^2)=Q^{\rm DIS}(x,t)$. This does not fix the scheme 
completely, but it is sufficient to fix it at 
the next-to-leading~$\ln x$ level. 
In particular, because $C_g^{2,\, {\rm\scriptstyle DIS}}=0$, eq.~\csc\ immediately
implies that for the DIS$\to$\MS\ scheme change
\eqn\distomsb{\eqalign{z^{qq}(N,\as(t))&=1-C_q^{2,\,\ms} (N,\as(t))\cr
z^{qg}(N,\as(t))&=- C_g^{2,\, \ms} (N,\as(t)).}}
Due to eq.~\llxsch\ and the fact that $\gamma_{gq}$ is only relevant
at the next-to-next-to leading~$\ln x$ level,  
the effect of the scheme change is thus fully
determined.

The small $x$ singular contributions to the large eigenvalue $\gamma^+$
 eq.~\sxexp\ are the same in the \MS\ and DIS schemes: indeed, eq.~\sxscheme\ implies
 that up to this order  the trace 
$\Tr \gamma=\gamma^++\gamma^- $ is invariant, but
 the singular part of $\gamma^-$ vanishes in both schemes, so
 $\gamma^+$ is invariant. Hence $\gamma^+_{ss}$ in eq.~\dlgampl\ is
 the same in the DIS and \MS\ scheme.

The function $\gamma^+_{ss}$ on which the resummation of ref.~\sxsym\
 is based, as  summarized in sect.~2, is however given in a scheme which differs
 from \MS\ by a small-$x$ scheme change eq.~\llxsch, the
 so--called \QMS\ scheme~\ciafqz. This is due to the fact that in the
 \MS\ scheme the running coupling corrections to duality
 eq.~\rccorrg, order by order in the expansion eq.~\sxexp\ in powers
 of $\as$ at fixed $\as/N$, are actually factorized in the coefficient
 function. These schemes are thus not suitable for the resummation
 eq.~\batandim\ of
the  running coupling corrections, because the singular contributions whose
 all--order resummation determines the leading small $x$ singularity sits 
in the coefficient function. The scheme--change function
 $u(M)$  which takes us from the \MS\ scheme to the \QMS\  scheme is of the 
form~\refs{\sxsym,\hefac,\sxap}
\eqn\rdef{u(M)=\frac{r(M)}{\sqrt{-\chi^\prime_0(M)}},}
where $\chi_0$ is the leading-order BFKL kernel eq.~\bfker\ and the
function $r(M)$ is regular in $0<{\rm Re}M<1$. Using eq.~\rdef\ in
 eq.~\schchgam\ one sees explicitly that the singular
 running--coupling contribution eq.~\rccorrg\ is removed from the
\QMS\ scheme anomalous dimension $\gamma_{ss}$ and factored in the
 coefficient function when transforming to the \MS\ scheme.

Clearly, in a consistent calculation the coefficient functions 
and the anomalous 
dimensions must all be computed in the same scheme. Here we have (at least) 
two choices: either we can work in the \QMS\ scheme throughout, or 
we can work in the \MS\ scheme.
If we choose to work in the \QMS\ scheme, using the \QMS\ scheme resummed anomalous
dimension $\gamma^+$ eq.~\rcsep, we must transform
the coefficient functions and quark--sector anomalous dimensions of
ref.~\ch\ from the \MS\ scheme by means of
the scheme change function eq.~\rdef. Alternatively, if we choose to 
work in the \MS\ (or DIS) scheme, we 
take the coefficient functions 
and quark sector anomalous dimensions from ref.~\ch, then we perform
the resummation of $\gamma^+$ in the \QMS\ scheme as discussed in
sect.~2.3, and finally we  transform $\gamma^+$ from 
\QMS\ scheme to \MS\ using \rdef\ and \schchgam. Because it is performed
after the resummation of $\gamma^+$, 
the scheme
change  results in subleading terms which multiply the resummed parton distribution, 
in the same way as a coefficient function, its
singularities, and in particular the singularity at $M=\half$, eq.~\rdef,  
may be treated in the same way as the pole in
eq.~\exrccf, i.e. removed by running coupling resummation. 

Let us discuss these two alternatives in detail. When working in \QMS\ scheme,
we start from 
$\gamma^{qg,\,\ms}_{ss}(\smallfrac{\as}{N})$ and  $c^{\ms}_1(N,M)$. We
then define~\summ\ the $Q_0$ version of the \MS\ scheme by constructing a
scheme change based on the requirements that
$\gamma^+$ be as in the \QMS\ scheme, while $\gamma^{qg}_{ss}$ is
unchanged, i.e. the same as in the \MS\ scheme. Assuming we perform
the scheme change through a leading~$\ln x$ scheme change eq.~\llxsch,\udef, 
followed by a next-to-leading scheme change eq.~\dlsch, the first requirement
fixes 
\eqn\umstoqz{z_{s}^{gg}\left(\smallfrac{\as}{N}\right)
=u\left(\gamma_s\left(\smallfrac{\as}{N}\right)\right),}
with $u(M)$ given by eq.~\rdef, and the second
requirement then, combining eq.~\schchgam\ with eq.~\llxsch\ and
demanding invariance of $\gamma^{qg}_{ss}$,
gives
\eqn\vmstoqz{z_{ss}^{qg}\left(\smallfrac{\as}{N}\right)
=\Big(1-\frac{1}{z_{s}^{gg}\left(\frac{\as}{N}\right)}\Big)
\frac{\gamma^{qg,\,\ms}_{ss}(\smallfrac{\as}{N})}{\gamma^{gg}_s(\smallfrac{\as}{N})}
\equiv v\left(\gamma_s\left(\smallfrac{\as}{N}\right) \right),}
where
\eqn\vdef{v(M)= \left(1-\frac{1}{u(M)}\right)\frac{h(M)}{M},}
with $h(M)$ defined in eq.~\hdef. Note that $v(M)$ is regular at $M=0$ since
$u(0)=1$. Consequently, given the series expansion for the 
coefficient functions $c_{ss}(M)$ (eq.~\conedef) in \MS, we can turn them into 
series expansions for the coefficient functions in \QMS\ by means of the combined 
scheme change eqs.~\umstoqz,\vmstoqz: the result is
\eqn\qzcss{\eqalign{c_{ss}^{2,g\,Q_0}(M)&= \frac{c_{ss}^{2,g\,\ms}(M)}{u(M)} 
-\left(1+\frac{h(M)}{M}\right)\left(1-\frac{1}{u(M)}\right),\cr
c_{ss}^{L,g\,Q_0}(M)&= \frac{c_{ss}^{L,g\,\ms}(M)}{u(M)}. }}
The quark coefficient functions are then given by the color-charge relation: 
modulo double counting terms,
\eqn\quarkcf{
c_{ss}^{2,q\,Q_0}=\smallfrac{C_F}{C_A}c_{ss}^{2,g\,Q_0};\quad
c_{ss}^{L,q\,Q_0}=\smallfrac{C_F}{C_A}c_{ss}^{L,g\,Q_0}.}

Constructing the \MS\ scheme is simpler in some respects, since the only scheme 
transformation necessary is that required to change $\gamma^+$ from \QMS\ to \MS, 
and this is given by \schchgam,\udef, and (the inverse of) \rdef. 
The 
main complication here is in the consistent treatment of running
coupling effects, which is necessary because $u(M)$ is
singular. The effect of this singularity arises
when $M$ acts on the parton distributions according to eq.~\iterglap,
so we can treat it with a  prescription based on that
which we used for the coefficient function in
eq.~\iterglaplo:
\eqn\iterglapdot{\eqalign{\smallfrac{d}{dt}[\gamma]&=\dot\gamma, \cr
\smallfrac{d}{dt}[\gamma^2]&=2\gamma\dot\gamma 
\left(1+\frac{\dot\gamma}{\gamma^2}\right), \cr
\smallfrac{d}{dt}[\gamma^3]&=3\gamma^2\dot\gamma 
\left(1+\frac{\dot\gamma}{\gamma^2}\right)
\left(1+2\frac{\dot\gamma}{\gamma^2}\right),\qquad{\rm etc.}\cr}}
so that
\eqn\gamplusmsbar{\eqalign{
\gamma^{+,\,\ms} &= \gamma^{+} 
- \smallfrac{d}{dt}\ln u\left([\gamma^{+\,res}_{\rm NLO}
\left(\as(t),N\right)]\right),\cr
&\qquad + \smallfrac{d}{dt}\ln u\big([\gamma^{+}_{0}(N)+\as
\gamma^{+}_{1}(N)-\smallfrac{\as}{N}\left(\smallfrac{n_c}{\pi}- \as
e_1^+\right)-e_2^+\smallfrac{\as^2}{N^2}]\big),\cr}}
where the second term is evaluated in practice by first expanding $u(\gamma)$ 
in powers of $\gamma$ and then using \iterglapdot. The last term is a 
matching term, chosen in just the same way as in eq.~\rescf\ to ensure 
that the resummation does not corrupt the anomalous dimension at large 
$N$ (and thus the splitting function at large $x$).

\subsec{Resummed quark anomalous dimensions}

Once the scheme is fixed, we can construct all the resummed anomalous 
dimensions and coefficient functions. 
The anomalous dimension 
$\gamma^+$ and its resummation in \QMS\ scheme are
as in ref.~\sxsym, as summarized in sect.~2 of the present paper.
If we are working in \MS, we must also add to this the contribution 
eq.~\gamplusmsbar, as explained above. 

The resummed quark sector anomalous dimensions are then built as
follows. First, we note that $\gamma^{qq}$ can be
obtained from  $\gamma^{qg}$ using the colour--charge relation
eq.~\colch. Furthermore, $\gamma^{gg}$ and $\gamma^{gq}$ can be
obtained from $\gamma^+$ and the quark-sector entries of $\gamma$
using eq.~\trdet. Hence we concentrate on the construction of $\gamma^{qg}$.
We start from $\gamma^{qg,\,\ms}_{ss}(\smallfrac{\as}{N})$, written 
as in eq.~\hdef.
We then construct the double--leading expression
\eqn\dlqg{\gamma^{qg}_{\rm DL}(\smallfrac{\as}{N})=\gamma^{qg}_{0}(N)+\as
\gamma^{qg,\,\ms}_{1}(N)
+\as \left(h\left(\gamma^{gg}_s(\smallfrac{\as}{N})\right)
-h(0)-h'(0)\smallfrac{n_c}{\pi}\smallfrac{\as}{N}\right),}
where the last two terms subtract the double counting.
We then note that the
function $h(M)$ eq.~\hdef\ has  singularities in $M$ analogous
to those of the contribution from  $c_{ss}(M)$ 
eq.~\conedef\ to the coefficient function: we  treat these by running
coupling resummation eq.~\iterglaplo. Putting everything together 
at the resummed
level we thus get
\eqn\resqg{\eqalign{\gamma^{qg}_{\rm res}(\smallfrac{\as}{N})
&=\gamma^{qg}_{0}(N)+\as
\gamma^{qg,\,\ms}_{1}(N)
+\as h\left([\gamma^{+\,res}_{\rm
NLO}\left(\as(t),N\right)]\right)\cr
&\quad - \as h\big([\gamma^{+}_{0}(N)+\as
\gamma^{+}_{1}(N)-\smallfrac{\as}{N}\left(\smallfrac{n_c}{\pi}- \as
e_1^+\right)-e_2^+\smallfrac{\as^2}{N^2}]\big)\cr
&\qquad\qquad -h(0)\as 
-h'(0)\left(\smallfrac{n_c}{\pi}- \as e_1^+\right)\smallfrac{\as}{N},}}
where by $h([\gamma])$ we mean that $h(M)$ is expanded in
power series of $M$, and then evaluated using eq.~\iterglaplo.
The terms in the last two lines are matching and double--counting
terms, which have been constructed  just as they were for the coefficient 
functions eq.~\rescf. 

This construction then gives us the 
resummed $\gamma^{qg}$ both in the
\MS\ and \QMS\ scheme. The resummed coefficient functions in either the 
\MS\ are then found using eq.~\rescf\ with the expression of $c_{ss}$ 
given in ref.~\ch, while its
counterpart in the  \QMS\ scheme is found in the same way, but with $c_{ss}$ given 
by the scheme change
eqs.~\qzcss, evaluated as function of
$[\gamma^{+\,res}_{\rm NLO}]$ in the sense of eq.~\iterglaplo.

\goodbreak
\newsec{Phenomenological Implications}
\subsec{Splitting functions and coefficient functions}

We now discuss the application of our results  and their phenomenological
 impact. We start by presenting the $n_f\not=0$  full  matrix of resummed 
singlet splitting functions together with the  coefficients $c^i_q$ and 
$c^i_g$ (i=2,L) for the singlet structure functions $F_2$ and $F_L$, 
respectively. The curves for the splitting functions for $\as=0.2$, 
$n_f=4$  (the average values relevant for HERA) versus $1/x$ are shown 
in figs.~1-2. There we show the plots of the gluon splitting functions 
$xP_{gg}$ and $xP_{gq}$ (fig.~1), and the quark splitting functions $xP_{qq}$ 
and $xP_{qg}$ (fig.~2) both in fixed order perturbation theory (at the LO, NLO 
and NNLO level) and in the resummed case (at LO and NLO). The NLO resummed 
curves are given both in the \QMS\ and in the \MS\ scheme, while at 
LO-resummed level there is no scheme difference. In fixed order 
perturbation theory there is no difference between $Q_0\MS$ and \MS\ 
up to and including the NNLO level but \QMS\ and \MS\ become 
different at higher orders. An important achievement of our work is 
the complete control of all aspects of scheme change at the resummed level.  
Resummed results for the splitting function matrix have also been 
obtained in ref.~\matevol\ but in a scheme which differs from the 
commonly used \MS\ scheme by  unknown, presumably small, terms. In 
contrast our results  are given in either the \QMS\ or the \MS\ 
schemes in a completely specified way.

\pageinsert
\vbox{
\vskip-0.5truecm
\hbox{\centerline{
\hskip1truecm
\epsfxsize=13.5truecm
\epsfbox{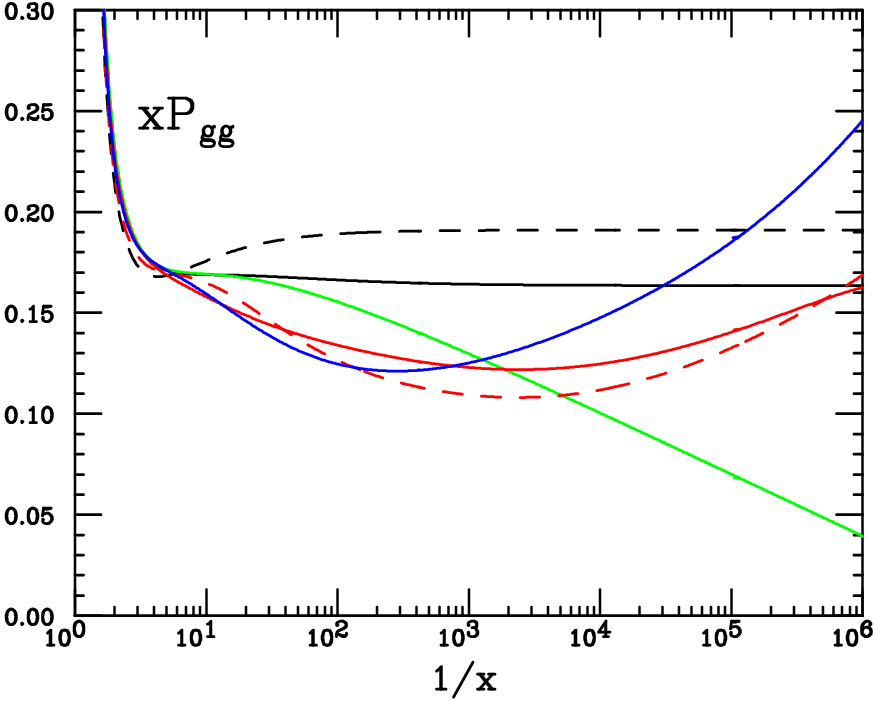}
}}
\hbox{\centerline{
\hskip1truecm
\epsfxsize=13.5truecm
\epsfbox{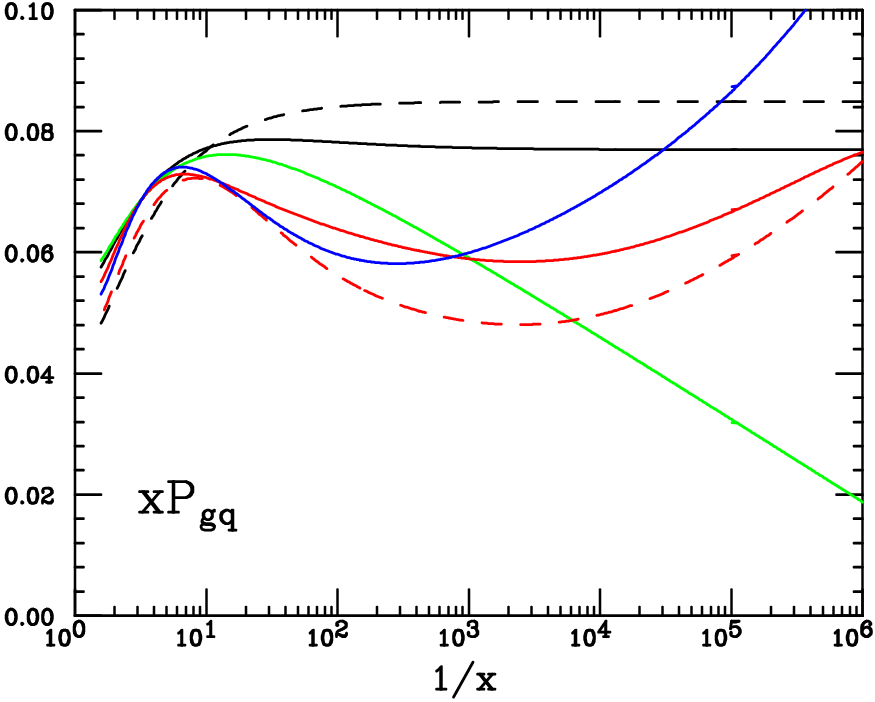}
}}
\hbox{
\vbox{\footnotefont\baselineskip6pt\narrower\noindent Figure 1: The 
gluon splitting functions $xP_{gg}$ and $P_{gq}$, plotted with $\as=0.2$ and 
$n_f=4$. The curves are (from top to bottom for $xP_{gg}$ at $x\sim0.2$): 
fixed order perturbation theory LO (black dashed), NLO (black  solid),  NNLO (green),
resummed LO (red dashed) and NLO in \QMS\ scheme  (red solid)
and in the \MS\ scheme (blue )
 }}\hskip1truecm}
\endinsert
\pageinsert
\vbox{
\vskip-0.5truecm
\hbox{\centerline{
\hskip1truecm
\epsfxsize=13.5truecm
\epsfbox{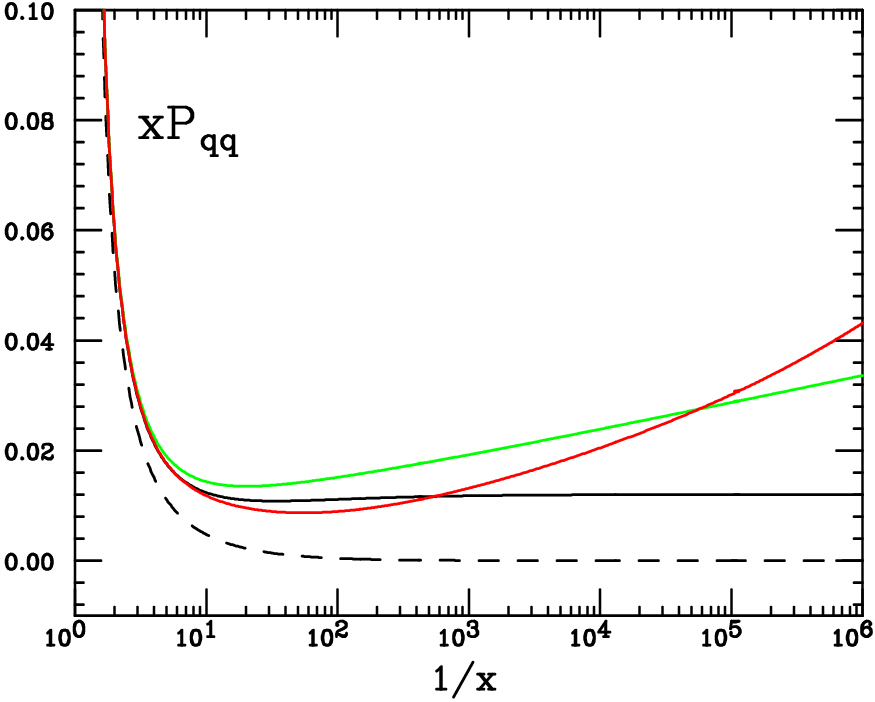}
}}
\hbox{\centerline{
\hskip1truecm
\epsfxsize=13.5truecm
\epsfbox{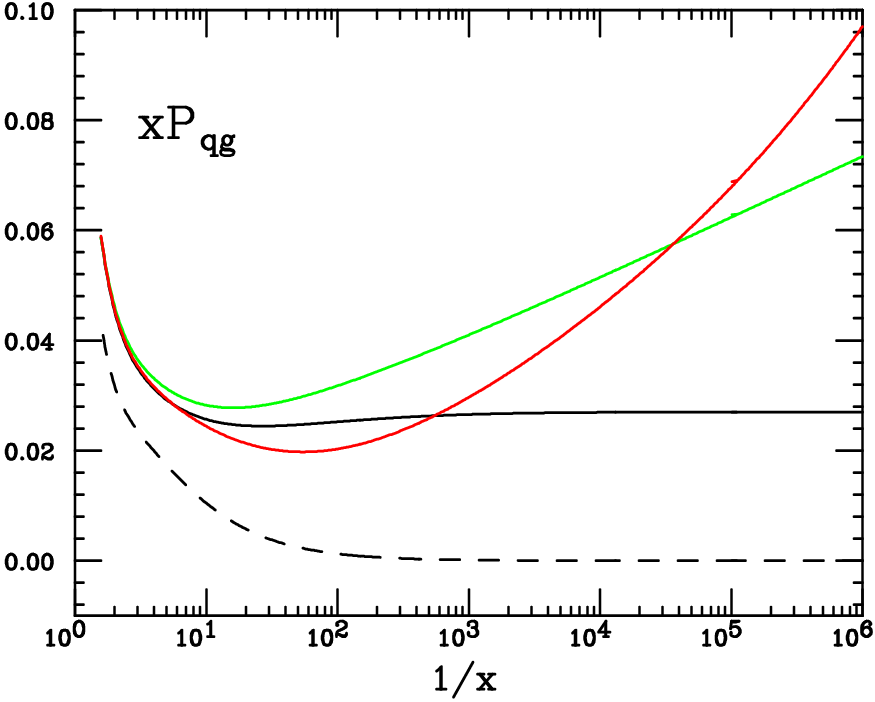}
}}
\hbox{
\vbox{\footnotefont\baselineskip6pt\narrower\noindent Figure 2: The 
quark splitting functions $xP_{qq}$ and $xP_{qg}$ plotted with 
$\as=0.2$ and $n_f=4$. The curves are as in fig.~1a (bottom to top for
$x\sim10^{-6}$). 
Note however that here the resummed LO coincides with
unresummed LO, and resummed NLO \MS\ coincides with resummed NLO \QMS.
 }}\hskip1truecm}
\endinsert
We now discuss the main features of the resulting splitting functions.  
We show the most important of the four, $xP_{gg}$,  in fig.~1. As 
for the $n_f=0$ case that we studied in our previous work \sxsym, 
the difference between the resummed and fixed order LO and NLO curves 
in the $x$ range relevant for collider experiments is moderate. But 
the strong instability shown by the NNLO fixed order perturbation 
theory is cured by the resummation. Above $x\sim 0.1$ the resummed 
curves match precisely the corresponding fixed order ones. For 
smaller values of $x$, the splitting function $xP_{gg}$ 
shows a significant dip directly inherited 
from $xP_{+}$~\ciafdip. 
This dip has important phenomenological consequences in 
that it extends the region of validity of the fixed order perturbative 
LO and NLO evolution. The onset of the truly asymptotic small $x$ rise, 
both in the resummed LO and the NLO (in the \QMS\ scheme) curves, is 
postponed to very small values of $x$. Note that for $xP_{gg}$ the \MS\ 
curves are considerably steeper than in the \QMS\ scheme at small $x$. 
This is the due to the singular nature of the scheme change eq.~\rdef\
which takes to \QMS, whose effects are 
only compensated when  the evolved parton densities are 
combined  with the coefficient functions. 

The splitting function $xP_{gq}$ is also plotted in fig.~1. It inherits 
most of the features of $xP_{gg}$, except that it is very soft at 
large $x$. Note that the absolute scale of  $xP_{gq}$ is about 
$\smallfrac{4}{9}$ 
smaller than that of $xP_{gg}$ at small $x$, which can be directly 
attributed to the ratio of colour Casimir factors $C_F/C_A$. 

The singlet quark splitting functions 
$P_{qq}$ and $P_{qg}$ are shown in fig.~2. Here the resummed LO 
coincides with the unresummed LO: indeed the resummation effects 
start at NLO. The NLO resummed curves at not too small values of 
$x$ are always bracketed between the LO and NNLO perturbative 
results. Moreover the singular parts of $xP_{qq}$ are again $\smallfrac{4}{9}$ 
times those of $xP_{qg}$ so that their resulting behaviour is similar.  
Recall also  that, as discussed in sect.~4.2, 
the \QMS\ scheme has been defined in such a way that 
$xP_{qq}$ and $xP_{qg}$  are the same as in \MS.  

\pageinsert
\vbox{
\vskip-0.5truecm
\hbox{\centerline{
\hskip1truecm
\epsfxsize=13.5truecm
\epsfbox{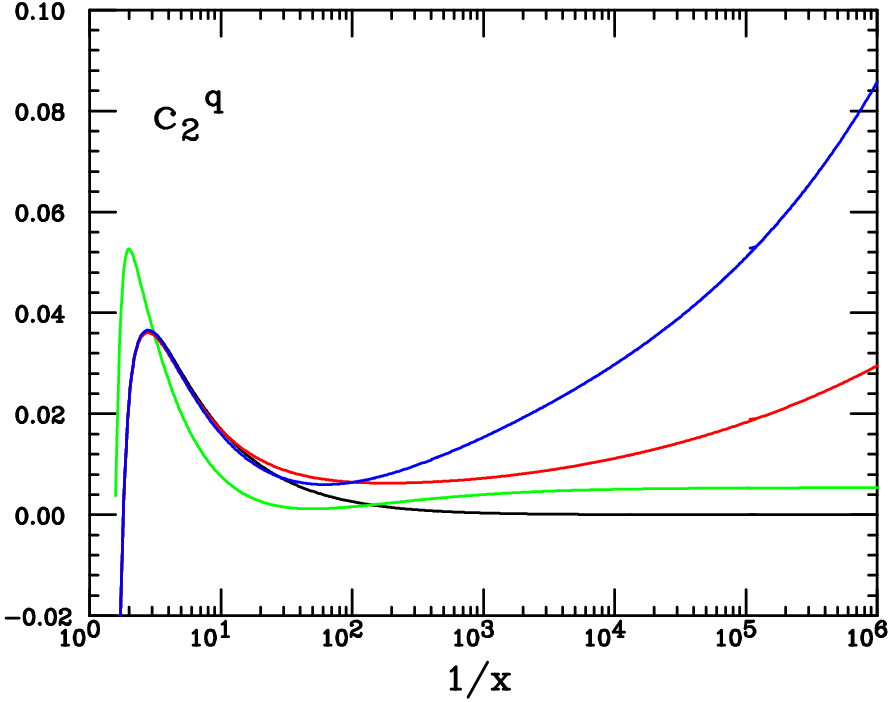}
}}
\hbox{\centerline{
\hskip1truecm
\epsfxsize=13.5truecm
\epsfbox{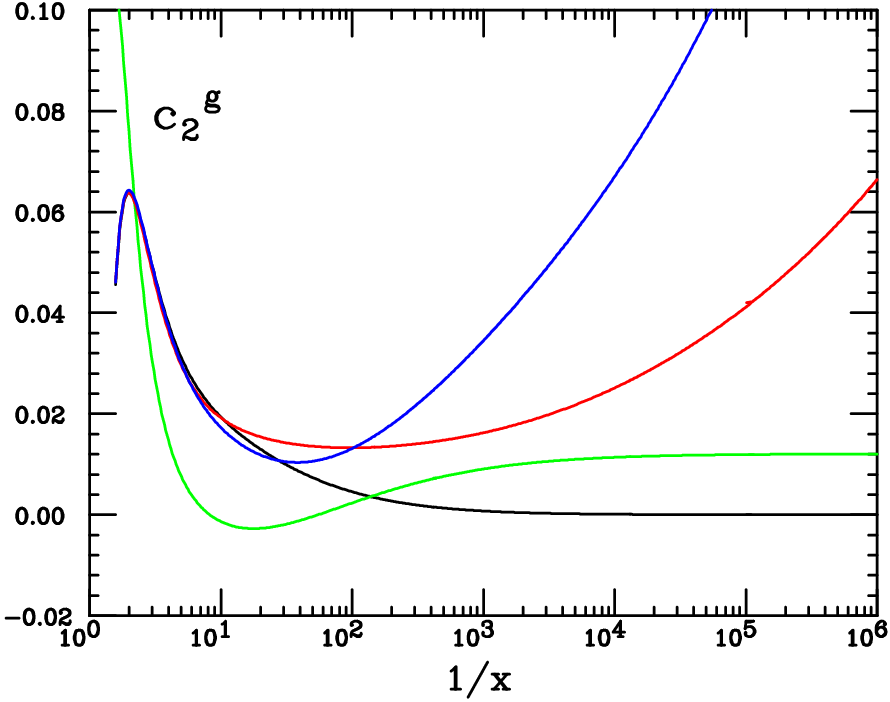}
}}
\hbox{
\vbox{\footnotefont\baselineskip6pt\narrower\noindent Figure 3: The 
$F_2$ coefficients plotted with $\as=0.2$ and $n_f=4$. The curves are
(bottom to top for 
$x\sim10^{-4}$): 
fixed order perturbative:  NLO (black solid), NNLO (green)
resummed NLO (red  solid) in \QMS\ scheme,
resummed NLO in the \MS\ scheme (blue).
 }}\hskip1truecm}
\endinsert
\pageinsert
\vbox{
\vskip-0.5truecm
\hbox{\centerline{
\hskip1truecm
\epsfxsize=13.5truecm
\epsfbox{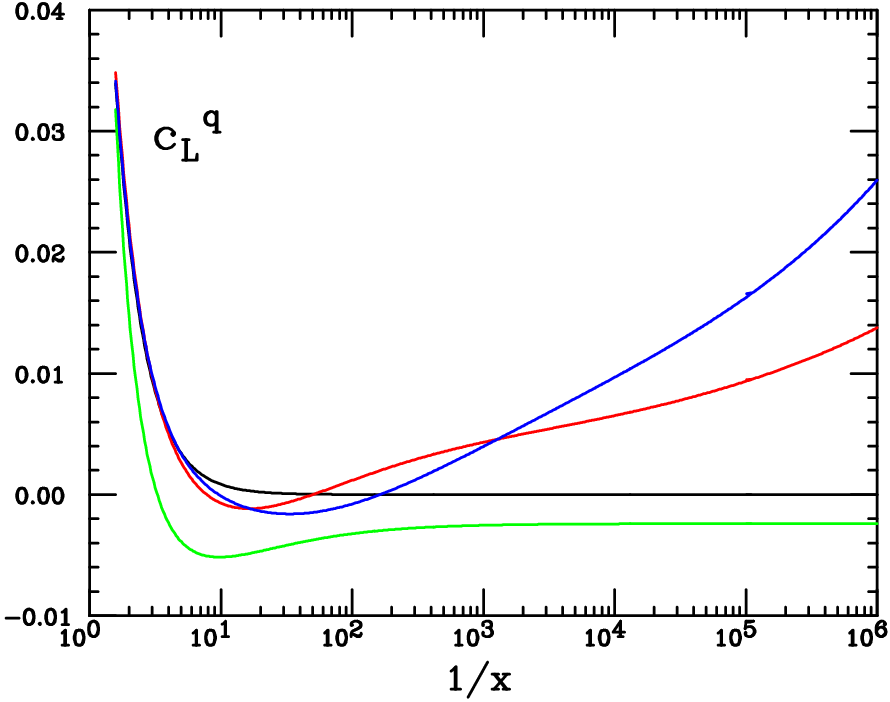}
}}
\hbox{\centerline{
\hskip1truecm
\epsfxsize=13.5truecm
\epsfbox{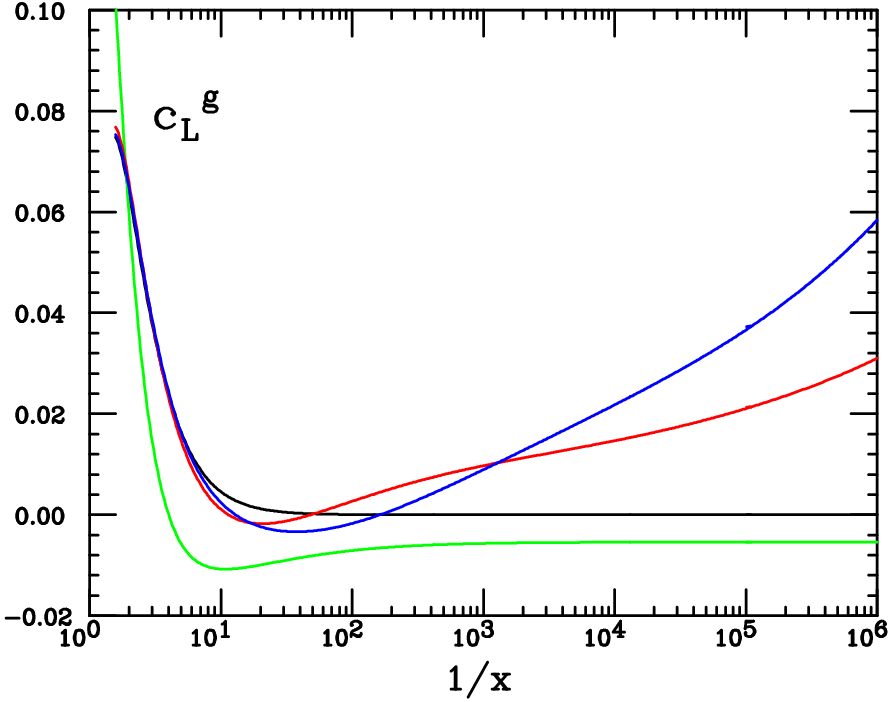}
}}
\hbox{
\vbox{\footnotefont\baselineskip6pt\narrower\noindent Figure 4: The
$F_L$ coefficients plotted with $\as=0.2$ and $n_f=4$. The curves are
(bottom to top for 
$x\sim10^{-6}$):  
fixed order perturbative:   NNLO (green), NLO (black),
resummed NLO (red) in \QMS\ scheme,
resummed NLO in the \MS\ scheme (blue).
 }}\hskip1truecm}
\endinsert

For the coefficient functions there are four curves in each of the 
$c^i_{q,g}$ plots in figs.~3-4. The curves refer to the NLO and NNLO 
fixed order perturbative results (the LO coefficients are either zero 
or, in the case of $c^2_q$, proportional to a delta function at $x=1$: 
$c^2_q=\smallfrac{5}{18} \delta (1-x)$) and to the NLO resummed coefficients in the 
\QMS\ and  \MS\ schemes. We see that also for the coefficient 
functions the \MS\ curves are steeper at small $x$ because in this
scheme the contributions corresponding to the singular term
eq.~\rccorrg\ are included in the coefficient function. 
For this reason, in our previous papers on the theory of resummed 
evolution for the singlet structure functions, we always adopted the 
\QMS\ scheme, where no unphysical singularities appear either in the 
splitting functions or in the coefficients. In a different scheme the 
combination of splitting functions and coefficients leads to a 
compensation in the evolution of physical quantities, which
end up  being different only through 
higher order effects, as we shall see later. 

\topinsert
\vbox{
\vskip-0.5truecm
\hbox{\centerline{
\hskip0.5truecm
\epsfxsize=8.5truecm
\epsfbox{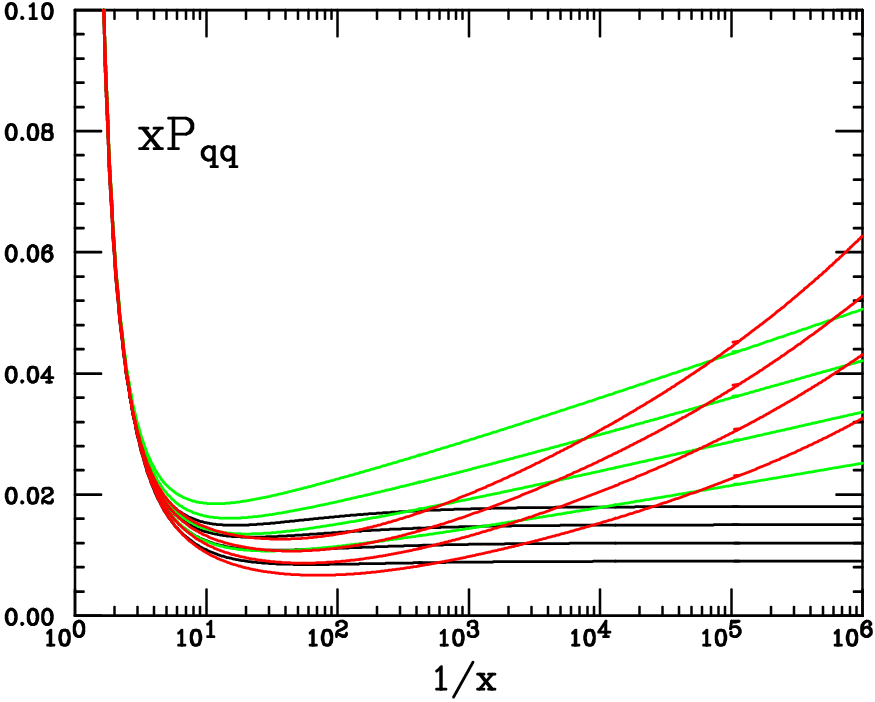}
\hskip-0.3truecm
\epsfxsize=7.8truecm
\epsfbox{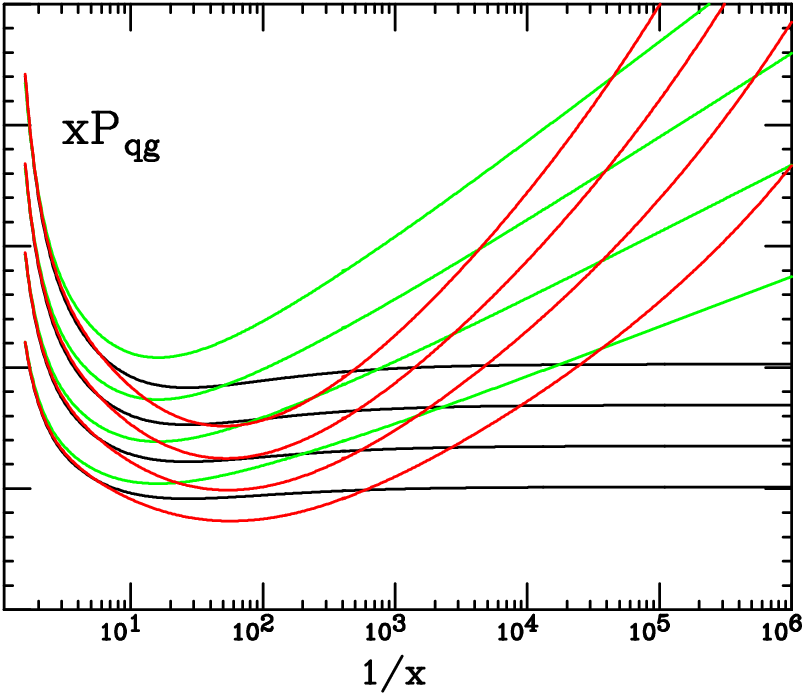}
}}
\vskip0.5truecm
\hbox{\centerline{
\hskip0.5truecm
\epsfxsize=8.5truecm
\epsfbox{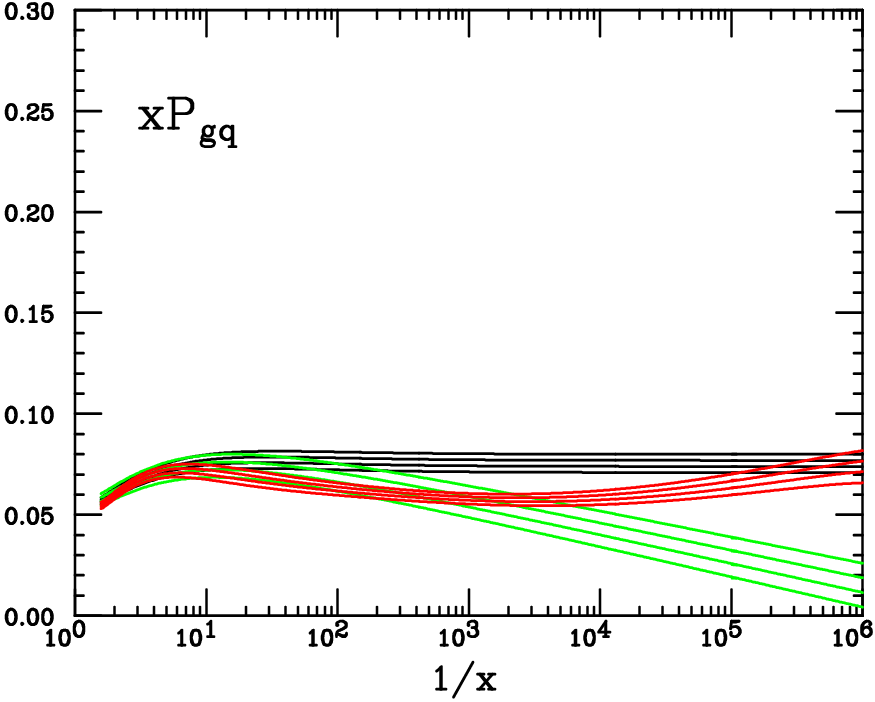}
\hskip-0.3truecm
\epsfxsize=7.8truecm
\epsfbox{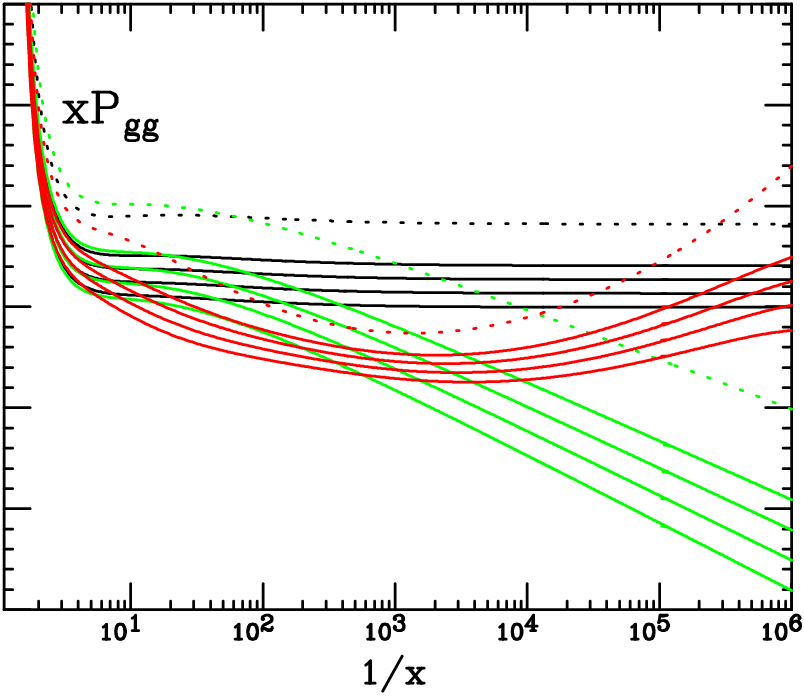}
}}
\vskip0.5truecm
\hbox{
\vbox{\footnotefont\baselineskip6pt\narrower\noindent Figure 5: 
$n_f$ dependence of splitting functions for $\as=0.2$. The plotted 
curves are for $n_f=3,4,5,6$ (solid), for $gg$ 
also $n_f=0$ (dotted) is shown for comparison. The three sets of 
splitting functions are: 
fixed order perturbative:  NLO (black), NNLO (green);
resummed NLO (red) in \QMS\ scheme. For $qq$ and $qg$ as $n_f$
increases the small $x$ value becomes larger: asymptotically constant
at NLO, stronger rise at NNLO and strongest rise at the resummed level.
For $gq$ and $gg$ as $n_f$
increases the small $x$ value becomes smaller: asymptotically constant
at NLO, stronger drop at NNLO and deeper dip at the resummed level.
 }}\hskip1truecm}
\endinsert
\topinsert
\vbox{
\vskip-0.5truecm
\hbox{\centerline{
\hskip0.5truecm
\epsfxsize=8.5truecm
\epsfbox{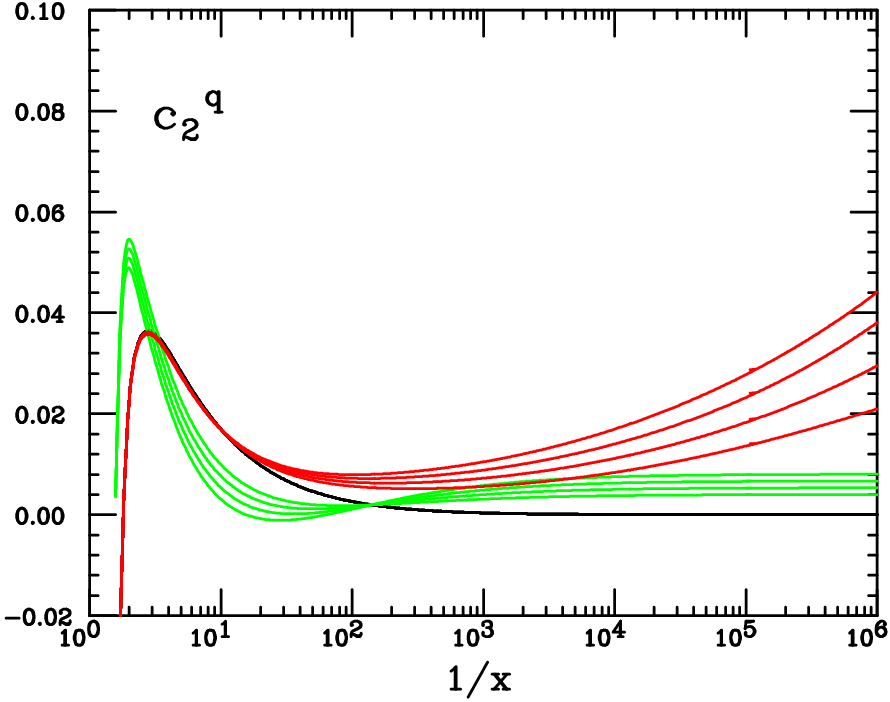}
\hskip-0.3truecm
\epsfxsize=7.7truecm
\epsfbox{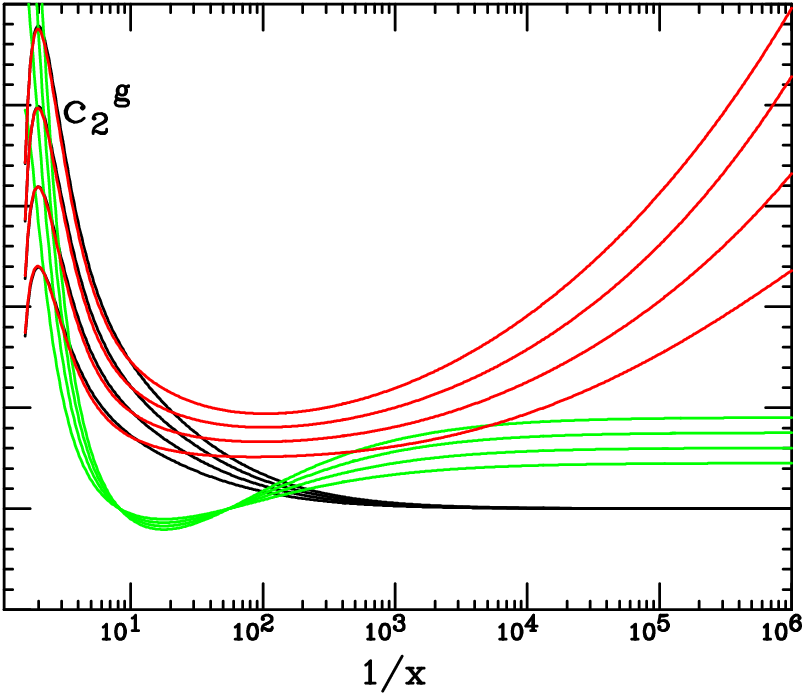}
}}
\vskip0.5truecm
\hbox{
\vbox{\footnotefont\baselineskip6pt\narrower\noindent Figure 6: The 
$n_f$ dependence of $F_2$ coefficients for $\as=0.2$. The plotted 
curves are for $n_f=3,4,5,6$ (solid).  The three sets of coefficients are: 
fixed order perturbative:  NLO (black), NNLO (green);
resummed NLO (red) in \QMS\ scheme.  
Note that the fixed order perturbative NLO  result 
is independent of $\as$. In all cases, for both $q$ and $g$, there is an 
increase with $n_f$ at small $x$.
 }}\hskip1truecm}
\endinsert
\topinsert
\vbox{
\hbox{\centerline{
\hskip0.5truecm
\epsfxsize=8.5truecm
\epsfbox{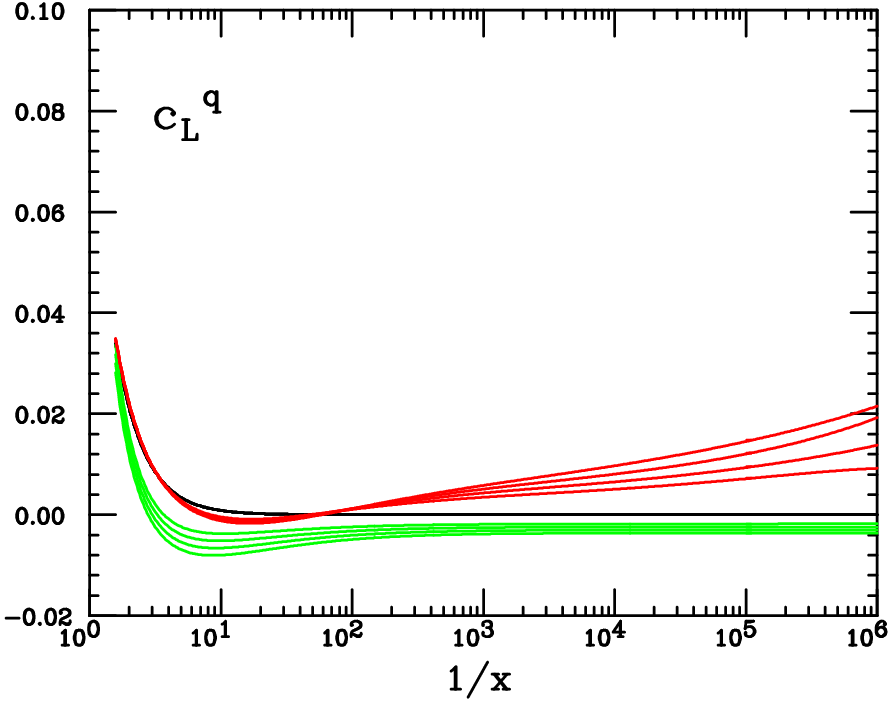}
\hskip-0.3truecm
\epsfxsize=7.7truecm
\epsfbox{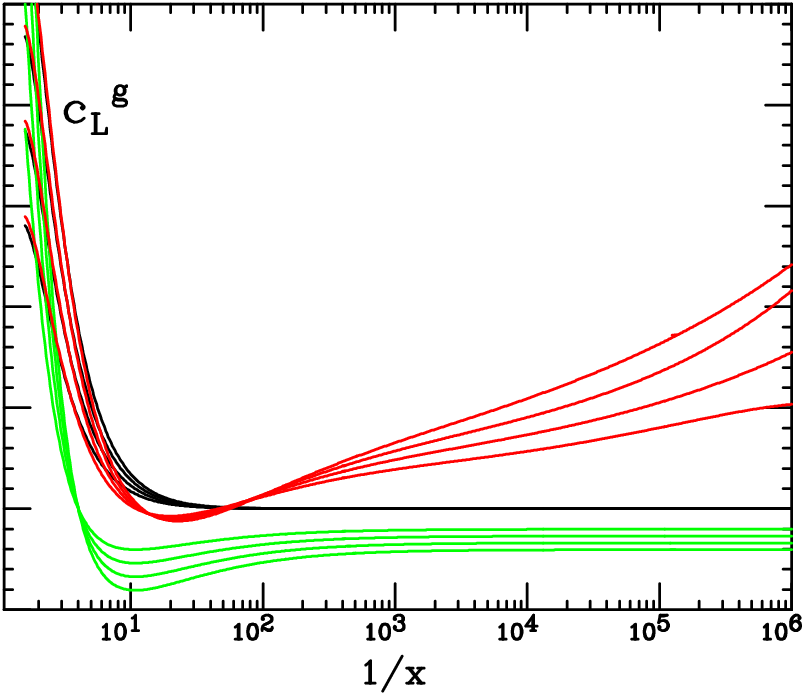}
}}
\vskip0.5truecm
\hbox{
\vbox{\footnotefont\baselineskip6pt\narrower\noindent Figure 7: As 
fig.~6, but now 
the $n_f$ dependence of $F_L$ coefficients. In all cases, for both 
$q$ and $g$, 
there is an increase in modulus with $n_f$ at small $x$, the resummed
result being positive while the fixed NNLO is negative.
}}\hskip1truecm}
\endinsert
In figs.~5-10 we show the separate dependence on $n_f$ and $\as$ of the 
splitting functions and coefficients. In practice, the relevant 
effective values of $n_f$ and of $\as$ are both functions of $Q^2$, 
so that their values move in a correlated way. However, for the 
sake of illustration, we show here the variation of one while the 
other is kept fixed. 

The $n_f$ dependence of  splitting functions and coefficients is 
displayed in figs 5-7. All these plots are for $\as = 0.2$. 
For clarity we only include the fixed order perturbative results at 
NLO and NNLO, and the resummed ones at NLO in the \QMS\ scheme.
The varying $n_f$ plots have $n_f= 3,4,5,6$  (solid) and $n_f= 0$  
(dotted) where relevant, for comparison.  The plotted curves  for 
the gluon splitting functions ($xP_{gg}$ and $xP_{gq}$) 
decrease as $n_f$ increases, 
while the
quark splitting functions ($xP_{qq}$ and $xP_{qg}$) increase. 
{}From these plots we see that varying $n_f$ does not make too much 
difference for $xP_{gg}$ and $xP_{gq}$, but the addition of more quarks softens 
the growth at small $x$. In the quark sector for splitting functions 
and for all the coefficient functions, there is an overall factor of 
$n_f$ in the main logarithmic terms. Thus the leading effect is an 
approximately linear rise in modulus with $n_f$ at small $x$; note
that the longitudinal NNLO coefficient functions are negative
at small $x$, and become more negative as $n_f$ increases.

\topinsert
\vbox{
\vskip-0.5truecm
\hbox{\centerline{
\hskip0.5truecm
\epsfxsize=8.5truecm
\epsfbox{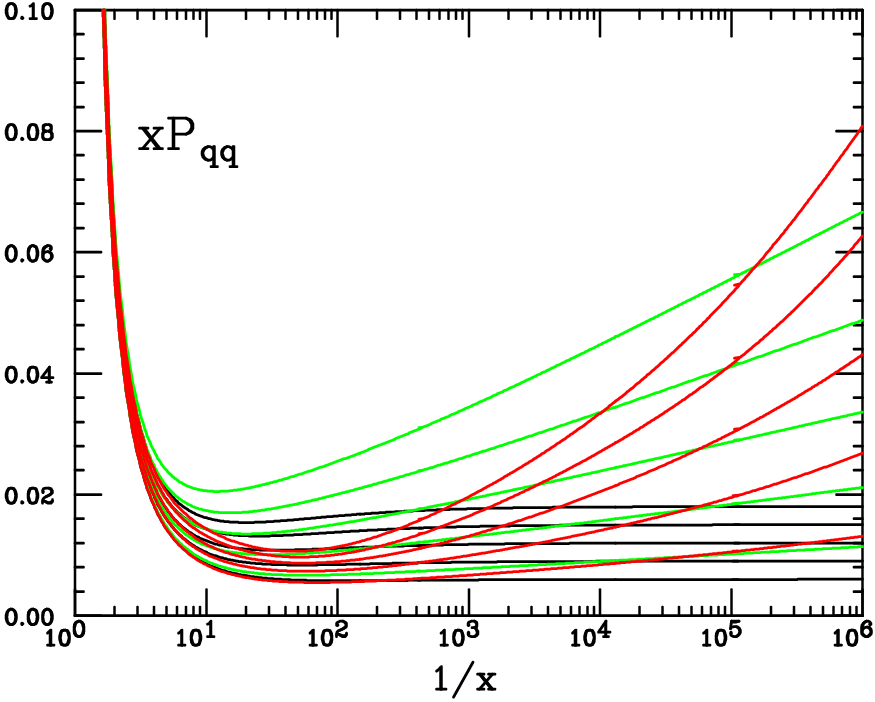}
\hskip-0.3truecm
\epsfxsize=7.8truecm
\epsfbox{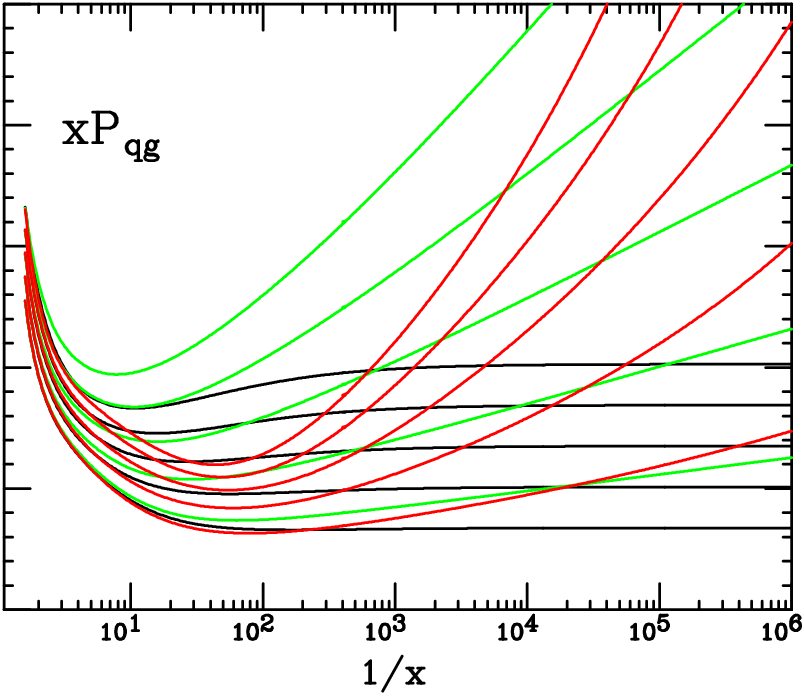}
}}
\vskip0.5truecm
\hbox{\centerline{
\hskip0.5truecm
\epsfxsize=8.5truecm
\epsfbox{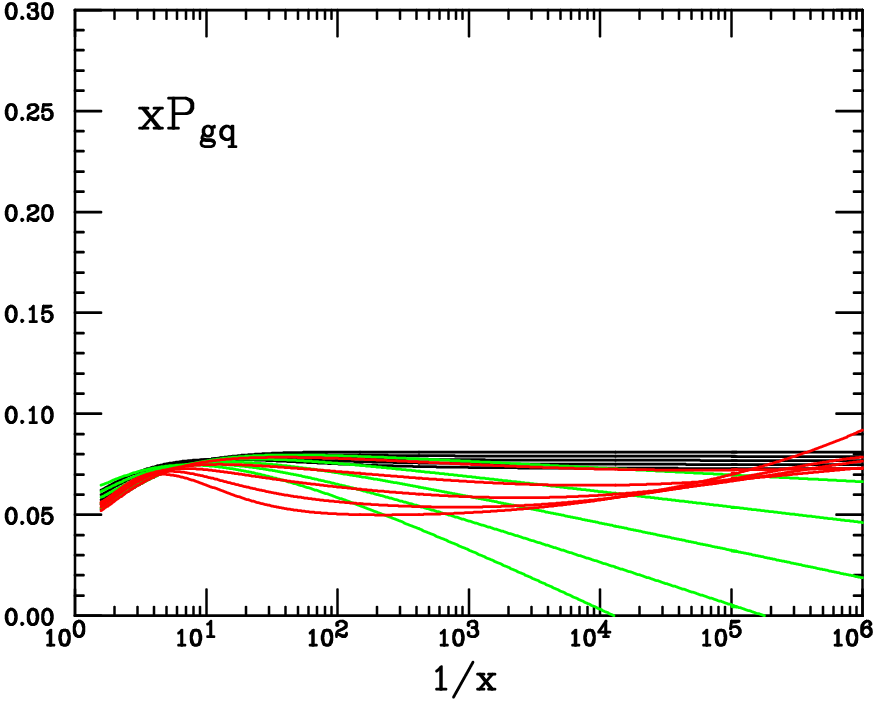}
\hskip-0.3truecm
\epsfxsize=7.8truecm
\epsfbox{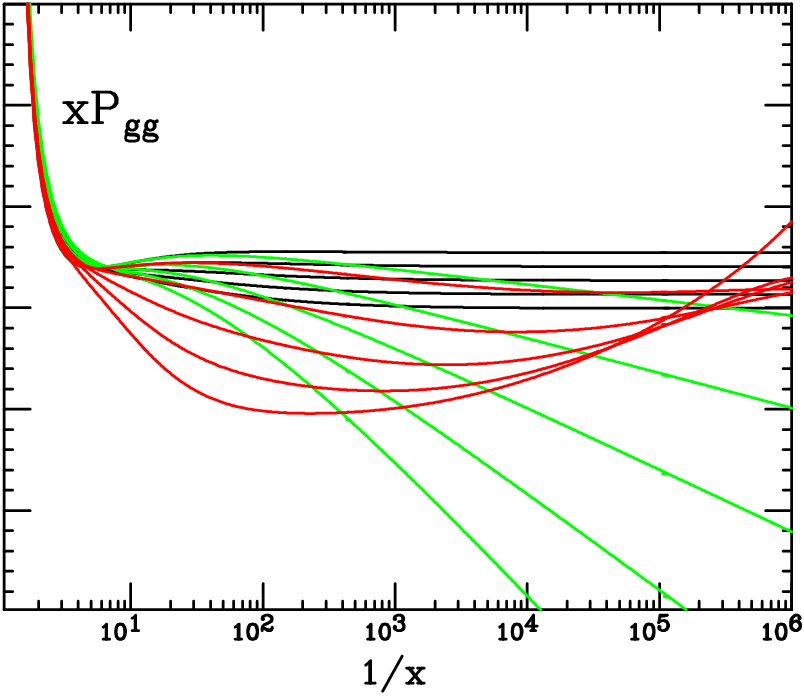}
}}
\vskip0.5truecm
\hbox{
\vbox{\footnotefont\baselineskip6pt\narrower\noindent Figure 8: 
$\as$ dependence of splitting functions for $n_f=4$. All curves 
are rescaled by $0.2/\as$, in order to eliminate the proportionality 
to $\as$. The values of $\as$ are  $\as=0.1, 0.15, 0.2, 0.25, 0.3$. 
The three sets 
of splitting functions are: 
fixed order perturbative:  NLO (black), NNLO (green);
resummed NLO (red) in \QMS\ scheme. At the resummed level, for $qq$ and $qg$ there is an increase with
$\as$ 
at small $x$, while for 
$gq$ and $gg$ a decrease at small $x$ and a deeper dip. At NNLO as
$\as$ increases there is 
a steeper rise for $qq$ and $qg$ and a steeper drop for $gq$ and $gg$.
 }}\hskip1truecm}
\endinsert

\topinsert
\vbox{
\vskip-0.5truecm
\hbox{\centerline{
\hskip0.5truecm
\epsfxsize=8.5truecm
\epsfbox{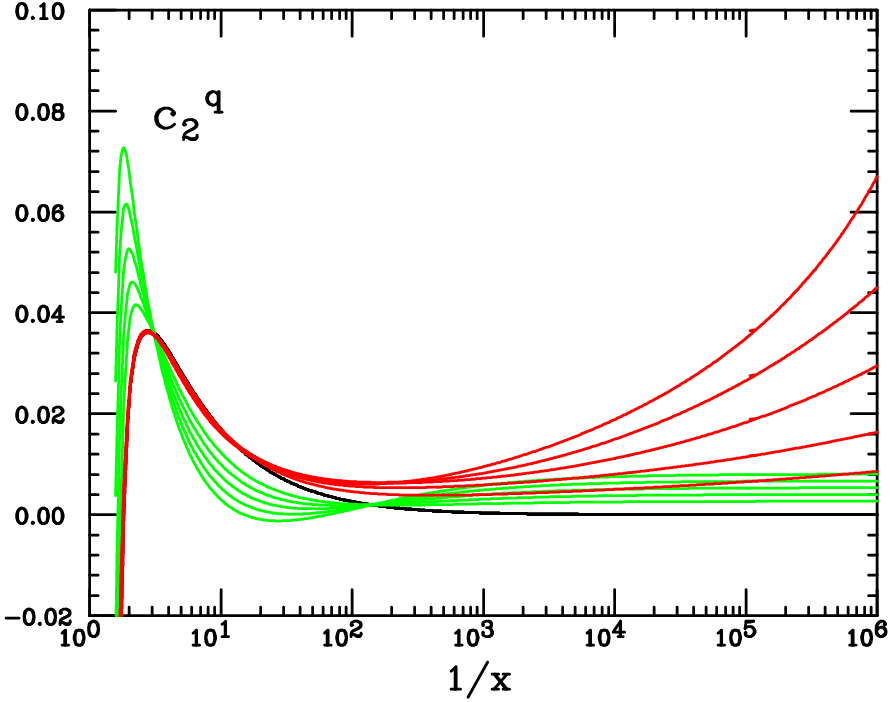}
\hskip-0.3truecm
\epsfxsize=7.7truecm
\epsfbox{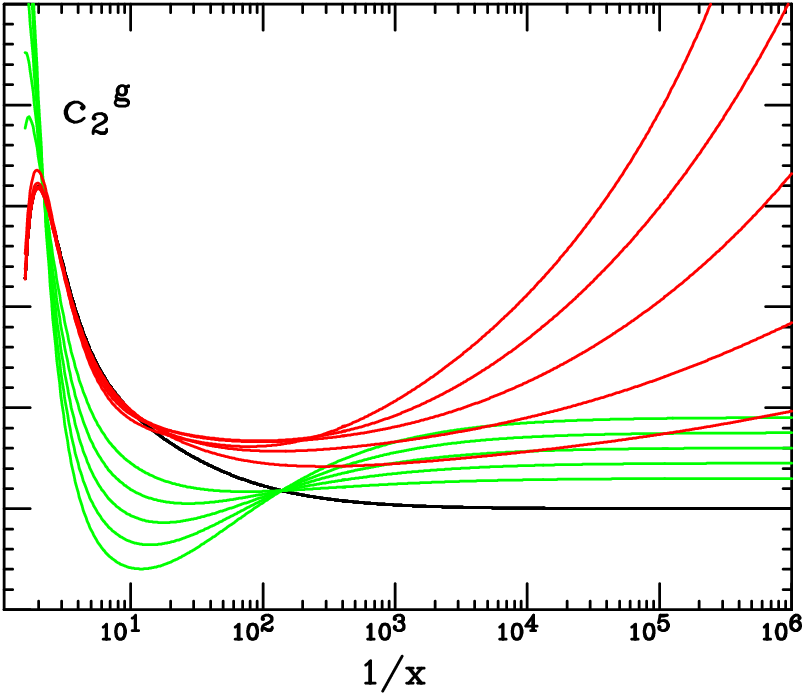}
}}
\vskip0.5truecm
\hbox{
\vbox{\footnotefont\baselineskip6pt\narrower\noindent Figure 9: The 
$\as$ dependence of $F_2$ coefficients.
All curves are rescaled by $0.2/\as$, in order to eliminate the 
proportionality to $\as$. The values of $\as$ are  
$\as=0.1, 0.15, 0.2, 0.25, 0.3$.  The three sets of coefficients are: 
fixed order perturbative:  NLO (black: note that the  result only 
depends on $n_f$ for $c_2^g$), NNLO (green);
resummed NLO (red) in \QMS\ scheme. 
For both coefficients there is in all
cases an increase 
with $\as$ at small $x$. The resummed curves rise at small $x$ while
the NLO and NNLO ones are asymptotically constant.
}}
\hskip1truecm}
\endinsert
\topinsert
\vbox{
\hbox{\centerline{
\hskip0.5truecm
\epsfxsize=8.5truecm
\epsfbox{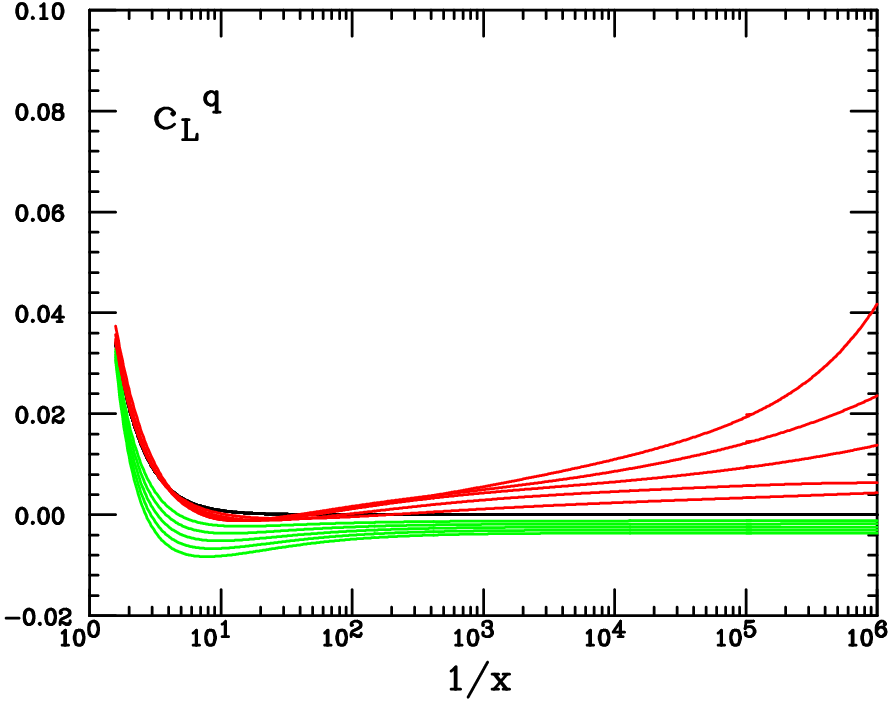}
\hskip-0.3truecm
\epsfxsize=7.7truecm
\epsfbox{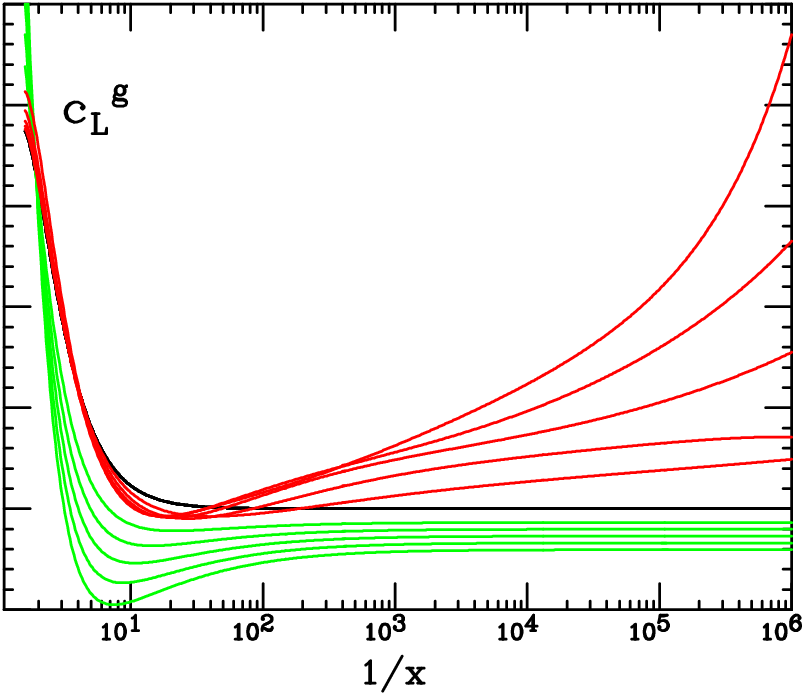}
}}
\vskip0.5truecm
\hbox{
\vbox{\footnotefont\baselineskip6pt\narrower\noindent Figure 10: As in fig.~9, but now 
the $\as$ dependence of the $F_L$ coefficients. 
For both coefficients at small $x$ there is an
increases with $\as$ in the resummed case (steeper growth) and a
decrease with $\as$ at NNLO (smaller negative asymptotic constant). 
 }}
\hskip1truecm}
\endinsert

The $\as$ dependences of  splitting functions and coefficients are 
displayed in figs 8-10. The varying $\as$ plots have 
$\as= 0.1,~0.15,~0.2,~0.25,~0.3$. All curves have $n_f=4$. 
In the plots the splitting functions and coefficients are 
multiplied by $0.2/\as$, so that the linear dependence on 
$\as$ is divided out, and the $\as=0.2$ curves are the same 
as those in figs 1-3. At low enough $x$, all the resummed splitting 
functions and coefficients show a steeper increase as $\as$ 
increases.  This is because the leading singularity in the 
$N$ plane moves further to the right making the asymptotic 
behaviour steeper at small $x$. At the same time $xP_{gg}/\as$ 
and $xP_{gq}/\as$ decrease in the intermediate region (i.e. the 
dip gets deeper). The rate of increase at small $x$ and the 
depth of the dip are related by a smooth interpolation between 
small and large $x$ and because of the integral constraint from 
momentum conservation (although the delta function terms at $x=1$ 
also depend on $\as$). The NNLO results, shown for comparison, as
$\as$ increases display a 
steeper rise at small $x$
of splitting functions in the quark sector and a steeper
drop in the gluon sector, and coefficient functions which, while
remaining asymptotically flat at small $x$, become larger in modulus
(positive for $c_2$ and negative for $c_L$)
Comparing results at the NNLO fixed order perturbative level with 
the resummed result at the NLO level, one sees that
the resummation improves the stability of the 
splitting functions even at rather large $\as$.

\pageinsert
\vbox{
\hbox{\centerline{
\hskip1truecm
\epsfxsize=14truecm
\epsfbox{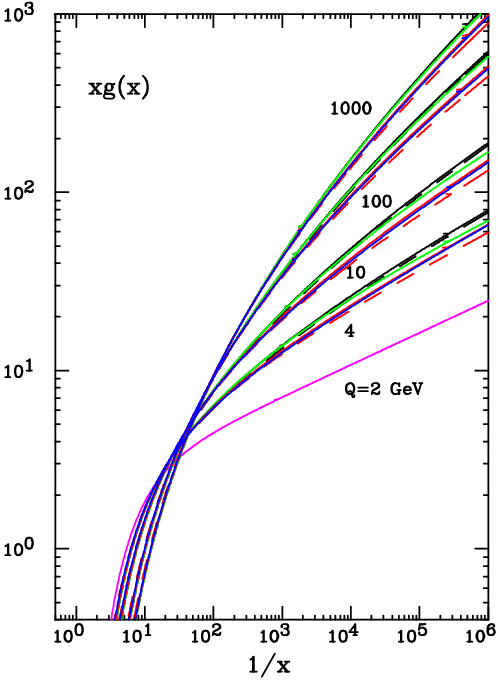}
}}
\hbox{
\vbox{\footnotefont\baselineskip6pt\narrower\noindent Figure 11: The 
small x behaviour of the  gluon distribution as a function of $1/x$ 
at different values of $Q=4,~10,~100,~1000$~GeV (for $\as=0.2$ and 
$n_f=4$). Also shown (purple) the initial parametrization at
$Q=2~GeV$. The curves are:
fixed order perturbation theory LO (black dashed), NLO (black solid), 
NNLO (green);
resummed LO (red dashed) and NLO (red solid) in \QMS\ scheme
resummed NLO (blue solid) in the \MS\ scheme. At all scales the fixed NLO
curve is highest (fixed LO slightly lower), 
the NNLO is lower, and the resummed NLO is lowest (resummed LO yet
slightly lower).  
Note that the blue and red curves (resummed NLO in the two schemes) are
almost indistinguishable.
 }}
\hskip1truecm}
\endinsert

\pageinsert
\vbox{
\hbox{\centerline{
\hskip1truecm
\epsfxsize=14truecm
\epsfbox{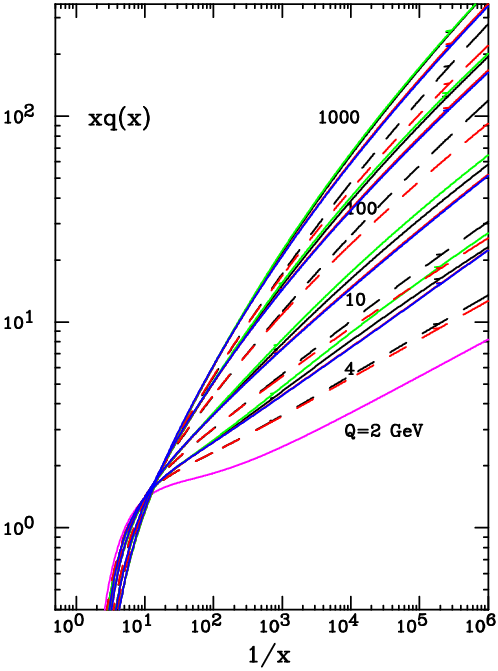}
}}
\hbox{
\vbox{\footnotefont\baselineskip6pt\narrower\noindent Figure 12: The 
small x behaviour of the total (valence plus sea) singlet quark 
distribution as function of $1/x$ at different values of 
$Q=4,~10,~100,~1000$~GeV (for $\as=0.2$ and $n_f=4$). Also 
shown (purple) the initial parametrization at $Q=2$~GeV.
The curves are: fixed order perturbation theory LO (black dashed), NLO 
(black solid), NNLO (green);
resummed LO (red dashed) and NLO (red solid) in \QMS\ scheme
resummed NLO (blue solid) in the \MS\ scheme. At all scales, the LO
curves are lowest (resummed below fixed order), 
resummed NLO higher, fixed NLO yet higher and fixed
NNLO highest.
Note that the blue and red curves (resummed NLO in the two schemes are
almost indistinguishable.
 }}
\hskip1truecm}
\endinsert
\subsec{Parton distributions and structure functions}

We now discuss the resulting evolution of parton densities and of 
structure functions and compare the resummed curves with the 
corresponding fixed order perturbative ones. In these plots $\as$ 
is running with $Q^2$ with $\as(m_{Z})=0.118$. The value of  $n_f$ is  
varied according to the zero--mass variable variable flavour number scheme:
contributions  from heavy quarks vanish below threshold and are
generated dynamically by perturbative evolution above threshold.  In 
figs.~11-12 we show the gluon and singlet quark parton densities 
as a function of $x$, down to $x=10^{-6}$, 
at different values of  $Q=2,~4,~10,~100,~1000$~GeV. The initial 
parton densities at $Q_0=2$~GeV are chosen to have a typical
simple semi-realistic shape,
adequate for our present purposes. Namely,  we take~\heralhc:
\eqn\gzdef{xg(x,t_0)=r_s x q_{sea}(x,t_0)= k_g x^{-0.18}(1-x)^5,}
\eqn\qzvdef{xq_v(x,t_0)= k_q x^{0.5}(1-x)^4,}
where $t_0=\ln Q_0^2/\mu^2$.
The constants $k_g$ and $k_q$ are fixed in such a way
 that the valence and momentum sum rules are satisfied. We choose
$r_s=3$,  so the fractions of momentum 
at $Q=Q_0$ are ~28$\%$ for valence quarks, ~18$\%$ for sea quarks 
and ~54$\%$ for gluons. 

The plots show, separately for the gluon 
and the total singlet quark densities, the initial distributions at 
$Q_0=2$~GeV, and then, for each higher value of $Q$, the 
resulting evolved distributions in the fixed order perturbative case,
 at LO, NLO and NNLO accuracy, and in the resummed case at LO and 
NLO (the latter both in the \QMS\ and \MS\ schemes). At small $x$ 
there is less evolution in the resummed cases than in fixed order 
perturbation theory. This is a consequence of the dip in the 
splitting functions.  In all cases the scheme dependence in the 
resummed NLO case is not very important.

We have also studied the renormalization scale dependence of the gluon and 
singlet quark evolution at fixed $x$ and $t$, by letting
\eqn\rensc{f(x,t;k)=f(x,t+\ln k)-\ln
k\frac{d}{dt}f(x,t+\ln k)+\half \ln^2
k\frac{d^2}{dt^2}f(x,t+\ln k),}
where $f(x,t)$ is the quark or gluon distribution.
At LO (fixed--order or resummed) 
only the first contribution on
the right--hand side is included; at NLO we include the first 
term evaluated at 
NLO, and the second evaluated at LO, while at NNLO we keep the first term 
evaluated at NNLO, the second evaluated at NLO and the third evaluated at LO.
It is easy to show that this is equivalent to the usual formulation of 
renormalisation scale variation where the argument of the running
coupling in the evolution equations is changed  
from $Q^2$ to $kQ^2$.

The scale  
variation eq.~\rensc\ provides an estimate of the size of the
theoretical uncertainty related to higher order contributions to
evolution equations, provided the expansion itself is uniform in $x$.
Hence, we expect it to provide a reliable estimate at the resummed
level, and to underestimate uncertainties at NLO and NNLO fixed order,
where scale variation cannot include the effect of higher--order
logarithmic terms. For example, in a NNLO fixed order calculation, 
scale variation explores the 
effect of subleading contributions to $xP_{gg}$ of the form $\as^4\ln x$, 
but fails to see the (known) $\as^4\ln^3 x$ contribution, which at 
small $x$ is of course much larger.

In fig.~13 we plot the evolved gluon density (starting from $Q_0=2$~GeV 
as described above) with $k$ eq.~\rensc\ 
varied between 0.1 and 10 at fixed $Q=10$~GeV 
and $x=10^{-2},~~10^{-4} ~{\rm or}~10^{-6}$. The scale dependence 
is larger at smaller values of $x$ where the $Q^2$ dependence is also 
steeper. The NLO approximations are much more stable against scale 
change than the LO counterparts, both for resummed and fixed order  
evolution, in agreement with expectations. The fixed order NNLO shows
comparatively little scale dependence, but as already mentioned this
does not include the effect of higher--order logs.

\pageinsert
\vbox{
\vskip-0.5truecm
\hbox{\centerline{
\epsfxsize=12.5truecm
\epsfbox{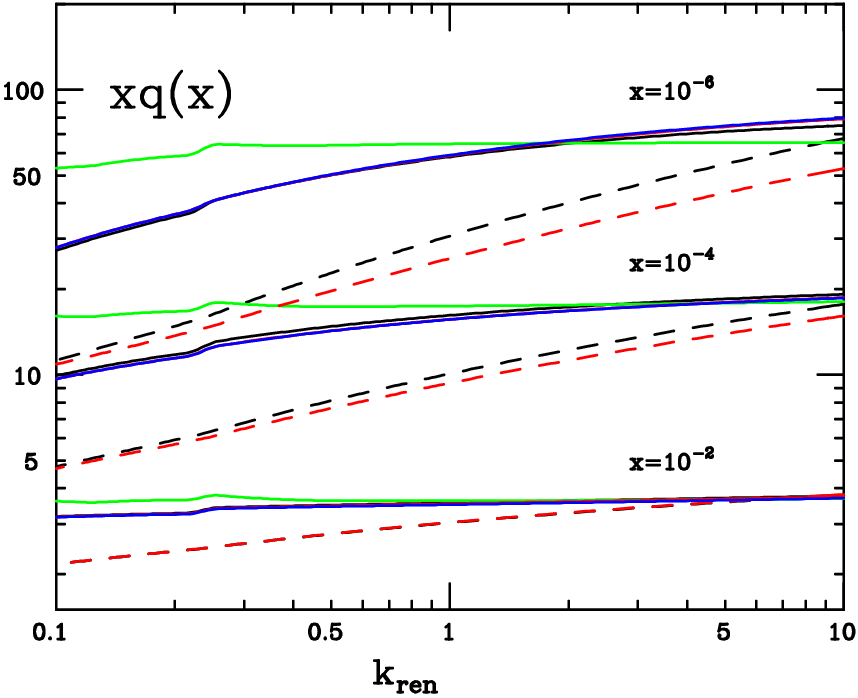}}}
\hbox{\centerline{
\epsfxsize=12.5truecm
\epsfbox{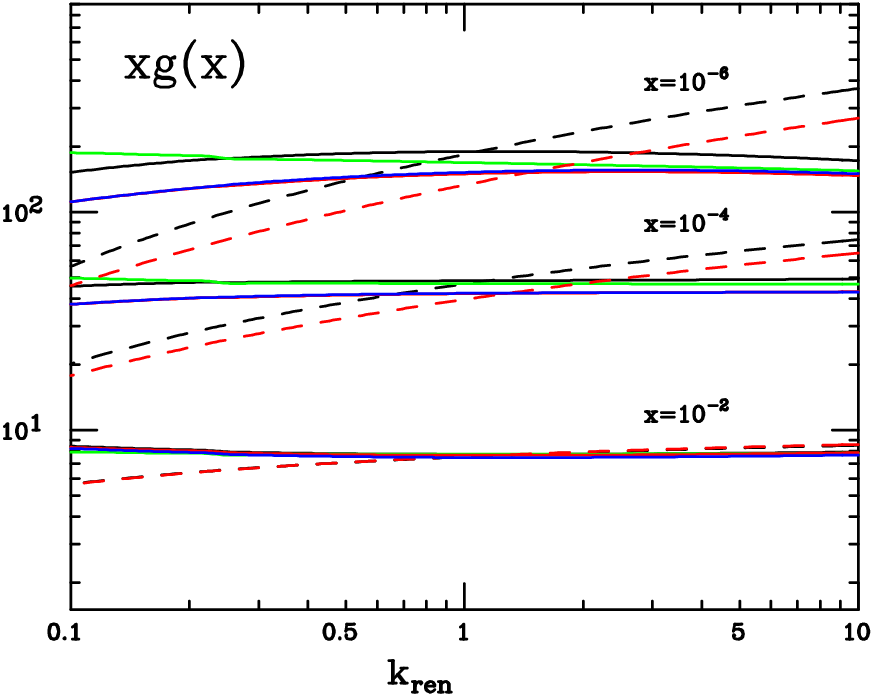}
}}
\vskip0.5truecm
\hbox{
\vbox{\footnotefont\baselineskip6pt\narrower\noindent Figure 13: The 
renormalization scale dependence eq.~\rensc\ of the singlet quark 
and of the gluon parton densities at fixed $Q=10$~GeV and $x=10^{-2}~ 
{\rm or}~10^{-4} ~{\rm or}~10^{-6}$ (with $n_f=4$). The curves are: 
fixed order 
perturbation theory LO (black dashed), NLO (black solid), NNLO ( green);
resummed LO (red dashed); resummed  NLO in \QMS\ scheme  (red solid)
and in \MS\ scheme  (blue solid). }}
\hskip1truecm}
\endinsert

Figures~11-12 explore the effects of various treatments of
perturbative evolution when the input parton distribution is kept
fixed. 
However, 
when higher--order corrections and
resummation are consistently included, the 
input physical observables are fixed and the parton distributions are
refitted.  Hence, a more realistic estimate of the
physical impact of the corrections can be obtained by assuming that
some physical observables are kept fixed (for example, the structure
functions $F_2$ and $F_L$) at a given scale, 
parton distributions are
determined from them, and then these parton distributions
are used to compute new physical observables (for example, 
structure functions or hadronic cross-sections at a higher scale). 

In practice, we proceed as follows.
We write the structure
functions $F(N,t)$  eq.~\strfmel\ as the product eq.~\strfundef\ of a matrix
coefficient function  $c(x,\as(t))$ eq.~\cfundef\ and an evolution
factor $\Gamma(N,t,t_0)$ eq.~\pdffac, times some initial parton
distribution $f(N,t_0)$:
\eqn\defobs{ F(N,t)= c(N,\as(t))\Gamma(N,t,t_0)f(N,t_0).
}
We then determine the structure functions at $Q_0=2$~GeV using the parton
distributions eq.~\gzdef-\qzvdef\ and the fixed order NLO expressions
of the coefficient functions:
\eqn\nloobs{ F_{\rm NLO}(N,t_0)= c_{\rm NLO}(N,\as(t_0))f(N,t_0).
}
Finally, we compute structure functions  with various  fixed--order and
resummed choices form the coefficient functions and anomalous
dimensions (and thus evolution factor), and with the input redefined
in such a way that at $Q_0^2$ all the structure functions are the same. 
So for instance at the resummed level we used resummed
expressions of the coefficient function and evolution factor, and take
as redefined input 
\eqn\redefin{ f_{\rm res}(N,t_0)= C^{-1}_{\rm
res}(N,\as(t_0))F_{\rm NLO}(N,t_0).
}
so that since 
\eqn\effres{F_{\rm res}(N,t)=C_{\rm res}(N,t)f_{\rm res}(N,t_0),}
$F_{\rm res}(N,t_0)=F_{\rm NLO}(N,t_0)$.

\pageinsert
\vbox{
\vskip-0.5truecm
\hbox{\centerline{
\epsfxsize=12.5truecm
\epsfbox{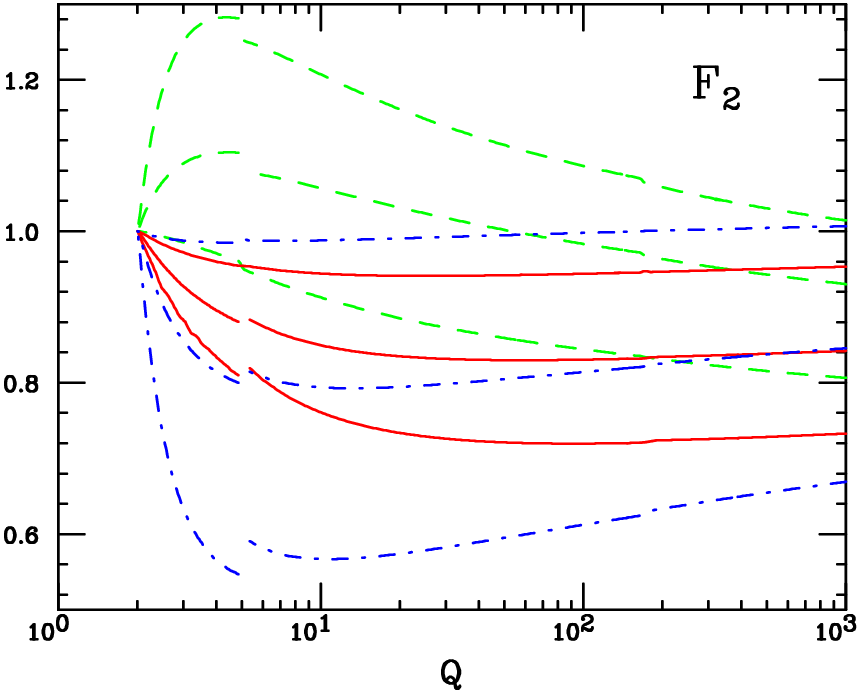}}}
\hbox{\centerline{
\epsfxsize=12.5truecm
\epsfbox{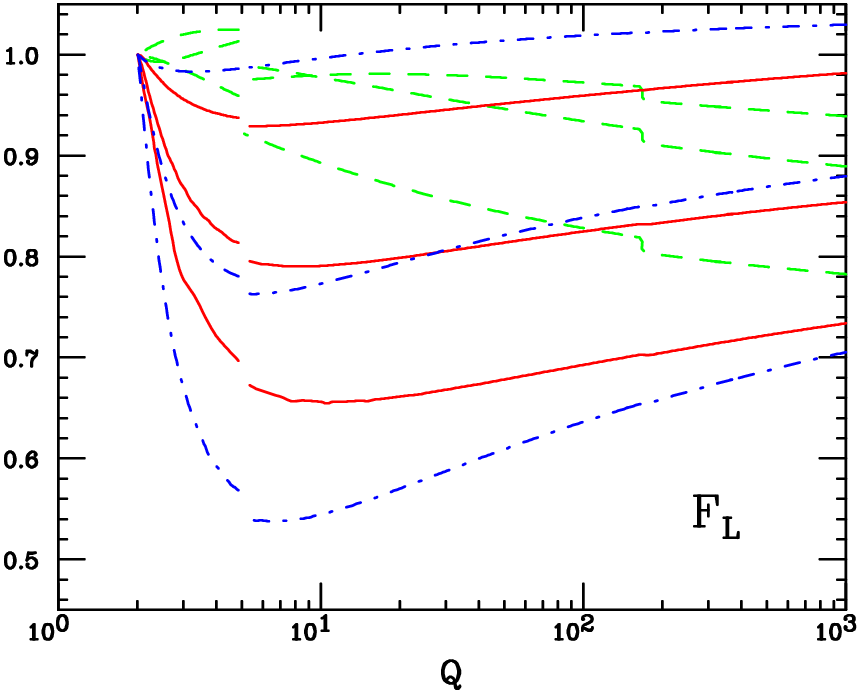}
}}
\vskip0.5truecm
\hbox{
\vbox{\footnotefont\baselineskip6pt\narrower\noindent Figure 14: The 
$K$-factors, defined as the ratio of the fixed order NNLO or resummed to the NLO
fixed order
 for the singlet $F_2$ and $F_L$ 
structure functions, with $F_2$ and $F_L$ kept fixed for all $x$ at
$Q_0=2$~GeV. Results are shown
at fixed  $x=10^{-2},~~10^{-4}~{\rm or}~10^{-6}$ as 
function of $Q$ in the range $Q=2-1000$~GeV with $\as$ running 
and $n_f$ varied  in a zero--mass variable flavour number scheme. 
The breaks in the curves correspond to the $b$ and 
$t$ quark thresholds. The curves are:   
fixed order perturbation theory NNLO (green, dashed);
resummed NLO  in \QMS\ scheme (red, solid),
resummed NLO in the \MS\ scheme (blue, dotdashed).}}
\hskip1truecm}
\endinsert
\pageinsert
\vbox{
\vskip-0.5truecm
\hbox{\centerline{
\epsfxsize=12.5truecm
\epsfbox{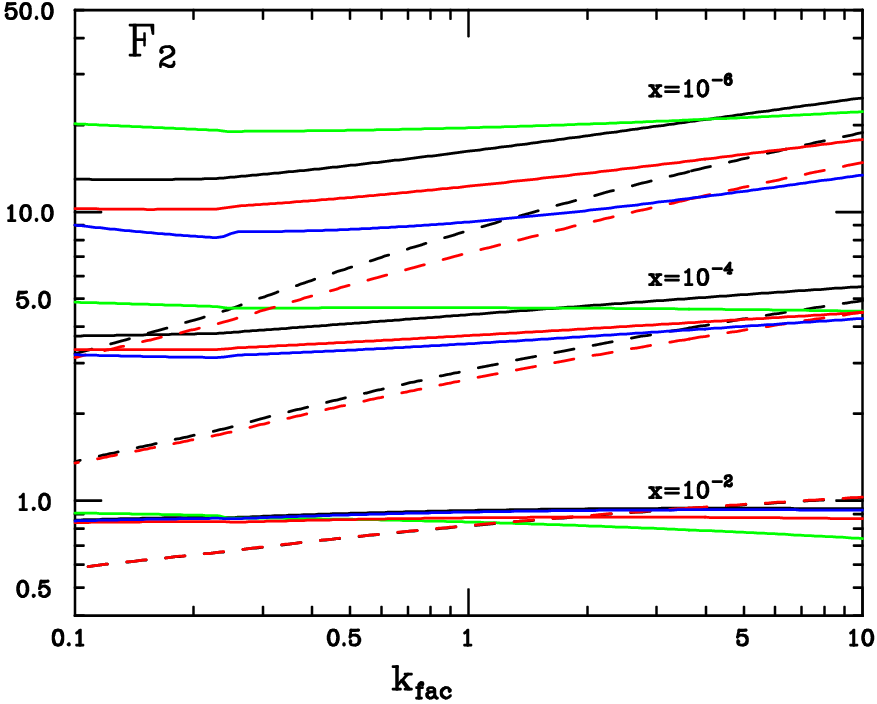}}}
\hbox{\centerline{
\epsfxsize=12.5truecm
\epsfbox{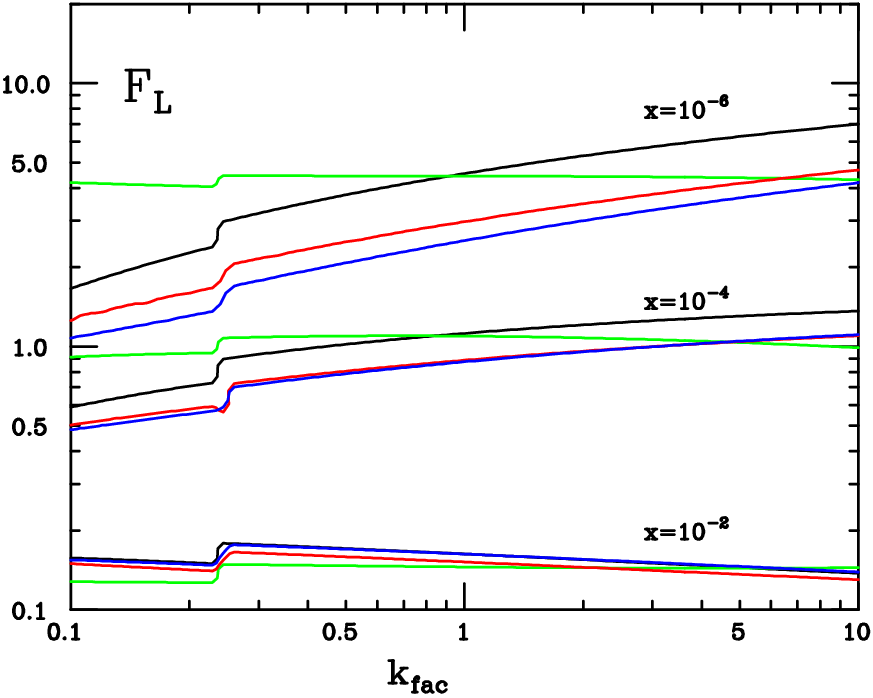}
}}
\vskip0.5truecm
\hbox{
\vbox{\footnotefont\baselineskip6pt\narrower\noindent Figure 15: The 
factorisation scale dependence eq.~(5.7) of the singlet $F_2$ 
and $F_L$ structure functions at fixed $Q=10$~GeV and $x=10^{-2},~~10^{-4} ~{\rm or}~10^{-6}$ (with $n_f=4$). The curves are: 
fixed order 
perturbation theory LO (black dashed, only $F_2$), NLO (black solid), 
NNLO ( green);
resummed LO (red dashed, only $F_2$) and NLO (red solid) in \QMS\ scheme
resummed NLO (blue solid) in the \MS\ scheme.
 }}\hskip1truecm}
\endinsert

In  fig.~14 we show  the $K$-factors
defined as the ratio of the NNLO fixed order or NLO resummed 
to the NLO fixed order
results for the singlet $F_2$ and 
$F_L$ structure functions, computed as a function of $x$ by Mellin
inversion of the expression eq.~\defobs\ with the input parton
distribution eq.~\redefin. We show results
 at fixed  
$x=10^{-2},~~10^{-4}~{\rm or}~10^{-6}$ 
as function of $Q$ in the range $Q=2-1000$~GeV, 
with $\as$ running and $n_f$ varied in a zero--mass variable flavour
number scheme as discussed above. The breaks in the curves 
correspond to the $b$ and $t$ quark thresholds and are a consequence of
the zero--mass approximation; in a more refined treatment they would
be replaced by a suitable matching at the heavy quark thresholds. 
For each $x$ value 
we present three curves: the resummed case in the \QMS\ scheme, 
the corresponding plot in the \MS\ scheme, and the NNLO fixed order 
perturbative.  The residual moderate scheme dependence between the 
\QMS\ and the \MS\ results, which is left after combining 
coefficients and parton densities, is only visible because of 
 the expanded linear scale of these 
plots, and is much smaller than  the scheme dependence of
coefficients and parton densities,  as was shown in figs.~2 and 3. 
It is interesting to observe that  
for $F_2$ the effect of resummation, at sufficiently small 
$x$ values, goes in the opposite direction to the NNLO perturbative 
evolution: the resummed $K$-factor is less than 1, corresponding to a 
smaller structure function at higher scales than with fixed order 
perturbative NLO evolution.

We have determined the factorization scale dependence of the structure
functions computed in this way, defined as
\eqn\factsc{F(x,t;k)=F(x,t+\ln k)-\ln
k\frac{d}{dt}F(x,t+\ln k)+\half \ln^2
k\frac{d^2}{dt^2}F(x,t+\ln k),}
where $F(x,t)$ is the structure function 
$F_2$ or $F_L$. Just as in eq.~\rensc, at 
LO only the first contribution on
the right--hand side is included; at NLO we include the first term evaluated 
at NLO, and the second evaluated at LO, while at NNLO we keep the first term 
evaluated at NNLO, the second evaluated at NLO and the third evaluated at LO.
It is easy to show that this procedure is equivalent to the usual one in which
the factorization scale is changed from $Q^2$ to $kQ^2$.
Results are shown in fig.~15. The general behaviour is similar as that
of renormalization scale variation in particular of for singlet quark
densities. However, $F_2$ displays 
less scale dependence than for quarks, thereby demonstrating the
cancellation of scheme dependence between coefficient function and
evolution factor. 
  The scheme dependence 
observed in the difference between the \QMS\ and \MS\ curves is 
consistent with the uncertainty obtained from the corresponding scale 
dependence.

Finally, in order to assess the variation in  parton
distributions when structure functions  are kept fixed, 
we have computed  $K$-factors for  $xq$ and $xg$
when these are determined by evolving up an input determined using 
eq.~\redefin. The $K$ factors are always defined as the ratio of
fixed order NNLO
or resummed to the NLO 
fixed order result. In fig.~16 we display these $K$-factors
at fixed  $x=10^{-4}~{\rm or}~
10^{-6}$ as function of $Q$ in the range $Q=2-1000$~GeV, with $\as$ 
running and $n_f$ varied, when $F_2$ and $F_L$ are fixed at three 
different reference
scales: $Q_0=2,~5,~10$~GeV (the corresponding curves can be identified 
by the point
where the plot starts). These plots compare the effects of evolution 
with and without resummation when the quark and gluon parton densities 
extracted at HERA, for example, at $x=10^{-4}$ and $Q_0=5$~GeV, are 
evolved at the same $x$ value up to, for example, $Q=100$~GeV 
for application to 
LHC phenomenology. 

\pageinsert
\vbox{
\vskip-0.5truecm
\hbox{\centerline{
\epsfxsize=12.5truecm
\epsfbox{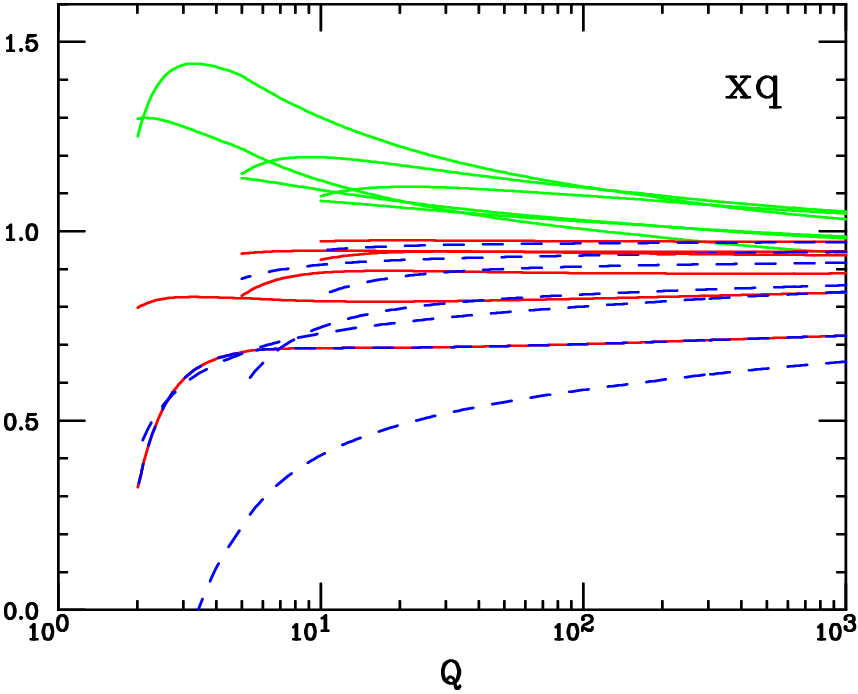}}}
\hbox{\centerline{
\epsfxsize=12.5truecm
\epsfbox{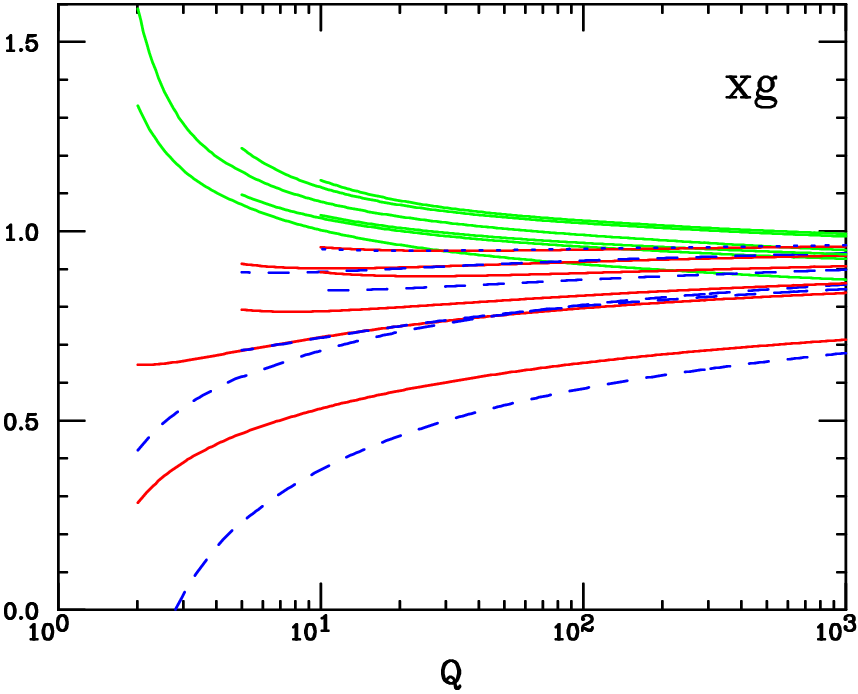}
}}
\vskip0.5truecm
\hbox{
\vbox{\footnotefont\baselineskip6pt\narrower\noindent Figure 16: The 
$K$-factors, defined as the ratio of the
fixed order NNLO or resummed to the NLO fixed order result, for  $xq$ 
and $xg$ at fixed  $x=10^{-4}~{\rm or}~10^{-6}$ as function of $Q$ 
in the range $Q=2-1000~GeV$, with $\as$ running and $n_f$ varied, 
when $F_2$ and $F_L$ are fixed, eq.~\redefin, 
at three different reference
scales: $Q_0=2,~5,~10~GeV$. The starting scale can be identified by the point
where the curves start.  The curves are: fixed order perturbation
theory NNLO (green, $K>1$ at low scale);
resummed NLO  in \QMS\ scheme (red),
resummed NLO in the \MS\ scheme (blue dashed).
}}\hskip1truecm}
\endinsert


Different important aspects can be observed from these plots. First, 
one can see the importance of resummation in extracting the parton 
densities from a given set of data at a given $Q$. It is well known 
that the data at smallest $x$ are obtained at smallest $Q$ on the 
average. Thus, for example, for $x=10^{-6}$ the most relevant curves 
would be those at $Q_0=2$~GeV. We see that, of course,
 from the same data on structure 
functions one extracts different values of the parton densities 
at the reference scale, depending on the perturbative order and
whether or not one resums.  Also, these different 
initial parton densities evolve in different ways. We can study this evolution,
using the appropriate splitting 
functions, up to the LHC domain, e.g at $Q=100$~GeV. We see that the 
evolution acts in the direction of decreasing the initial differences. 
In fact, if we had the data at the LHC and we could extract the parton 
densities directly at $Q=100$~GeV, the spread would necessarily be much 
smaller, because $\as(t)$ is smaller, and the resummation effects would 
be less important. Thus the evolution, in a consistent way, tends to reduce 
the differences obtained at smaller values of $Q$. The general conclusion 
is that,  if resummation effects are disregarded, the associated error in 
evolving the parton densities from HERA to the LHC is of the order of 
$5-20\%$. Note that the NNLO corrections and the resummed ones go in 
opposite directions, thus amplifying their difference. Of course, for 
the computation of physical quantities then also the hard partonic 
cross-section must be resummed and this would lead to an enhancement 
which partially compensates for the parton density suppression.
Finally, fig.~16 clearly shows that  the scheme 
dependence is moderate, while the resummation effects are rapidly 
becoming large at small $x$.

\vfill\eject

\newsec{Conclusion}

We have by now achieved a good understanding of the behaviour of singlet 
structure functions and of their evolution at small $x$ well beyond the 
domain explored at HERA and down to very small values of $x$, for all 
values of $Q^2$ where the leading twist approximation is applicable. 

The failure of the BFKL expansion on the one side and the remarkably 
accurate description of the HERA data by fixed order perturbation 
theory on the other, demanded a theoretical explanation.  The 
calculation of three loop splitting functions (NNLO)  has made the 
problem more acute along the way, by showing that at higher orders 
the fixed order expansion does indeed start to diverge. The 
solution of the problem is by now well established: 
important formally subleading terms cannot 
be ignored and must also be resummed. Their effect is to push the onset of
the  
truly asymptotic regime down to smaller values of $x$, while in an 
intermediate region, relevant for HERA data, the evolution shows a 
shallow dip, somewhat lower than the NLO perturbative result. The 
resummation procedure is determined by solid guiding principles from 
duality, momentum conservation, symmetry under gluon exchange of the 
BFKL kernel and accurate implementation of running coupling effects. 

In the present paper, after a summary of the general theoretical 
principles and methods,  we have developed the formalism to the level 
needed for a direct application to the calculation of physical observables. 
Resummed  
splitting functions and coefficients have been evaluated at general 
values of $n_f$ and $\as$ and their scheme dependence has been 
studied in detail, 
together with a check that this dependence is drastically reduced when 
the different ingredients are collected together to form physically 
measurable quantities. Finally we have studied the size of the resulting 
effects for predictions in the LHC region, showing that their 
importance while not dramatic is however certainly sizeable, and as
big as that of fixed order NNLO terms.

\bigskip\noindent
{\bf Acknowledgements}
\smallskip
Two of us (G.A. and S.F.) acknowledge partial supported from the 
Italian Ministero dell'Universit\`a e della 
Ricerca Scientifica, under the PRIN program for 2007-08. The work of R.D.B.  
has been done in the context of the Scottish Universities' Physics Alliance.
This work was partly supported by the Marie Curie
Research and Training network HEPTOOLS under contract
MRTN-CT-2006-035505.
We thank M.~Ciafaloni, D.~Colferai and G.~Salam for useful discussions.

\footatend\vfill\supereject\immediate\closeout\rfile\writestoppt
\baselineskip=14pt\centerline{{\bf References}}\bigskip{\frenchspacing%
\parindent=20pt\escapechar=` \input refs.tmp\vfill\eject}\nonfrenchspacing

\vfill\eject
\bye